\documentclass[5p,times,sort&compress]{elsarticle}

\usepackage[utf8]{inputenc}
\usepackage[american]{babel}
\usepackage{amsmath,braket,bbm,amssymb,xspace,graphicx}
\usepackage[version=3]{mhchem}
\usepackage{siunitx}
\usepackage[colorinlistoftodos,
            textsize=small,
            disable,
            ]{todonotes}
\usepackage[breaklinks,colorlinks]{hyperref}
\usepackage[protrusion=true,
            expansion=true,
            auto=true,
            kerning=true,
            spacing=true,
            tracking=false]
            {microtype}

\graphicspath{{fig_paper/}}
\journal{\ldots}

\newcommand*\tdContent[2][]{%
  \def\empty{}
  \def\arg{#1}
  \ifx\arg\empty
    \todo[inline,color=orange!30]{#2}
  \else
    \todo[inline,color=orange!30,caption={#1}]{#2}
  \fi
}
\newcommand*\tdRef[2][]{%
  \def\empty{}
  \def\arg{#1}
  \ifx\arg\empty
    \todo[inline,color=green!30]{#2}
  \else
    \todo[inline,color=green!30, caption={#1}]{#2}
  \fi
}

\newcommand*\veps  {\varepsilon}
\newcommand*\HH    {\mathcal{H}}
\newcommand*\MM    {\mathcal{S}}
\newcommand*\TT    {\mathcal{T}}
\newcommand*\UU    {\mathcal{U}}
\newcommand*\VV    {\mathcal{V}}
\renewcommand*\AA  {\mathcal{A}}
\newcommand*\KK    {\mathcal{K}}
\newcommand*\RR    {\mathcal{R}}
\newcommand*\C     {{\mathbb C}}
\newcommand*\R     {{\mathbb R}}
\newcommand*\one   {\mathbbm{1}}
\newcommand*\bk    {{\ensuremath{\vec k}}\xspace}
\newcommand*\br    {\vec{r}}
\newcommand*\ktilde{\vec{q}}
\newcommand*\tk    {k\xspace}
\newcommand*\bR    {{\ensuremath{\vec R}}}
\newcommand*\bZ    {{\vec 0}}
\newcommand*\W     {\text{w}}
\newcommand*\U     {\text{u}}      
\newcommand*\me    {m_\text{e}}
\newcommand*\eg    {$\text{e}_g$\xspace}
\newcommand*\ttg   {$\text{t}_{2g}$\xspace}
\newcommand*\ii    {\text{i}}
\newcommand*\ee    {\text{e}}
\newcommand*\intk  {\mathop{\int\limits_\text{\bz}\!     d\bk}}
\newcommand*\intw  {\mathop{\int_{-\infty}^{+\infty}\!\! d\omega}}
\newcommand*\intkw {\mathop{
    \int\limits_\text{\bz}\!d\bk \! \int_{-\infty}^{+\infty}\!\!d\omega
  }}

\DeclareMathOperator{\Tr}{Tr}
\renewcommand*\vec[1]{\boldsymbol{#1}}

\newcommand*\wien    {\textsc{Wien}2k\xspace}
\newcommand*\wtow    {\textsc{wien2wannier}\xspace}
\newcommand*\wannier {Wannier90\xspace}
\newcommand*\woptic  {woptic\xspace}
\newcommand*\Woptic  {Woptic\xspace}

\newcommand*\code[1]{\texttt{#1}}

\newcommand*\lapwi {\code{lapw1}\xspace}
\newcommand*\optic {\code{optic}\xspace}
\newcommand*\woprog{\code{woptic}\xspace}
\newcommand*\womain{\code{woptic\_main}\xspace}
\newcommand*\reftet{\code{refine\_tetra}\xspace}
\newcommand*\convvr{\code{convert\_vr}\xspace}

\newcommand*\Minterp{\code{interp}\xspace}
\newcommand*\Moptic {\code{optic}\xspace}
\newcommand*\MInterp{\code{Interp}\xspace}
\newcommand*\MOptic {\code{Optic}\xspace}

\newcommand*\abbrev[1]{\MakeUppercase{#1}}
\newcommand*\Abbrev[1]{
  \expandafter\newcommand\csname#1\endcsname{\abbrev{#1}\xspace}
}

\Abbrev{dmft}
\Abbrev{bz}
\Abbrev{pbe}
\Abbrev{gga}
\Abbrev{wf}
\Abbrev{vsc}
\Abbrev{dfg}
\Abbrev{fwf}
\Abbrev{erc}
\Abbrev{start}
\Abbrev{id}
\Abbrev{sfb}

\newcommand*\wfs {\abbrev{wf}s\xspace}
\newcommand*\ldad{\abbrev{dft+dmft}\xspace}
\newcommand*\lda {\abbrev{dft}\xspace}

\newcommand*\svo{\ce{SrVO3}\xspace}
\newcommand*\al{\ce{Al}\xspace}

\let\term\textit

\begin{document}

\begin{frontmatter}
  \title{\woptic: optical conductivity with Wannier functions and
    adaptive \tk-mesh refinement}

  \author[vienna,graz]{E. Assmann\corref{cor}}
  \cortext[cor]{Corresponding author}
  \ead{elias.assmann@tugraz.at}
  \author[vienna] {P. Wissgott}
  \author[prague] {J. Kuneš}
  \author[vienna] {A. Toschi}
  \author[vienna2]{P. Blaha}
  \author[vienna] {K. Held}

  \address[vienna]{Institute of Solid State Physics, TU Wien, 1040
    Vienna, Austria}%
  \address[graz]{Institute of Theoretical and Computational Physics,
    Graz University of Technology, 8010 Graz, Austria}%
  \address[prague]{Institute of Physics, Academy of Sciences of the
    Czech Republic, 18221 Praha 8, Czechia}%
  \address[vienna2]{Institute of Materials Chemistry, TU Wien, 1060
    Vienna, Austria}%

  \begin{abstract}
    We present an algorithm for the adaptive tetrahedral integration
    over the Brillouin zone of crystalline materials, and apply it to
    compute the optical conductivity, dc conductivity, and
    thermopower.  For these quantities, whose contributions are often
    localized in small portions of the Brillouin zone, adaptive
    integration is especially relevant.  Our implementation, the
    \textit{\woptic} package, is tied into the \wtow framework and
    allows including a many-body self energy, e.g. from dynamical
    mean-field theory (\dmft).  Wannier functions and dipole matrix
    elements are computed with the \lda package \wien and \wannier.
    For illustration, we show \lda results for fcc-\al and \dmft
    results for the correlated metal \svo.
  \end{abstract}

\end{frontmatter} 
 
\tdRef[referees]{Must submit the names and institutional e-mail
  addresses at least 3 potential referees.}

\listoftodos

\section{Introduction} 

The theoretical description of crystalline solids is greatly
simplified by their periodicity.  The Bloch theorem for
non-inter\-acting electrons allows one to replace a sum over
infinitely many discrete lattice vectors by an integral over a
continuous \tk-vector restricted to the Brillouin zone (\bz).  A
similar simplification is possible in interacting systems where the
crystal momentum \bk is a conserved quantity for a variety of
excitations.  This makes \bz-integration an indispensable part of any
numerical technique for periodic solids.  To evaluate the integral
numerically, we must discretize the \bz in some way.  The usual
methods used in band-structure calculations rely on an \textit{a
  priori} choice of the \tk-mesh which covers the \bz uniformly;
e.g. using a straightforward summation \cite{monkhorst_special_1976,
  chadi_special_1973, chadi_special_1977}, or the more sophisticated
\term{tetrahedron method} \cite{blochl_improved_1994}.  This is a
natural choice for the calculation of quantities such as the charge
density, to which all \tk-points contribute.  On the other hand, the
transport or low-energy spectral properties are usually dominated by
certain regions of the \bz, e.g. the vicinity of the Fermi surface.
To compute such quantities, an inhomogeneous \tk-mesh adapted to a
particular material may be a better choice.

In the present article, we describe a technique to recursively
generate an inhomogeneous \tk-mesh for periodic solids in three
dimensions.  Our implementation, the \textit{\woptic} package, is
designed to calculate the optical conductivity, dc conductivity, and
thermopower of interacting electrons.  However, the adaptive \tk-mesh
management is encapsulated in a subprogram (\reftet) which may easily
be adapted to other quantities.

\Woptic operates in the context of dynamical mean-field theory (\dmft)
\cite{metzner_correlated_1989, georges_hubbard_1992} for real
materials.  This ``\ldad'' approach \cite{kotliar_electronic_2006,
  held_electronic_2007} uses the band structure from
density-functional theory (\lda) in the local-density or generalized
gradient approximations (\gga) to construct an effective multiband
Hubbard model, which is analyzed using the \dmft technique.%
\footnote{The calculations reported in this article use the \gga in
  the form of the \pbe functional \cite{perdew_generalized_1996}.} %
Calculation of the optical conductivity represents a post-processing
step in this scheme.  In the present implementation, we use inputs
generated by the \wien \cite{blaha_full-potential_1990}, \wtow
\cite{kunes_wien2wannier:_2010}, and \wannier
\cite{mostofi_wannier90:_2008} codes and a self energy (on the
real-$\omega$ axis) from any \dmft solver.  So far, only local self
energies $\Sigma(\omega)$ are implemented, but the approach allows
including any self energy on top of \wien.  In particular, extension
to a non-local $\Sigma(\bk, \omega)$ (e.g. from $GW$
\cite{hedin_new_1965, tomczak_combined_2012}, or from extensions of
\dmft \cite{toschi_dynamical_2007, rubtsov_dual_2008,
  maier_quantum_2005}) is simple as long as $\Sigma(\bk, \omega)$ may
be obtained at any $\bk$.

When vertex corrections are neglected (they are strongly suppressed in
\dmft, see Sec.~\ref{Problem}), the optical conductivity involves a
\bz sum of contributions obtained from the \tk-resolved one-particle
spectral functions and dipole matrix elements between the
corresponding wave functions.  We start by evaluating the optical
conductivity on a uniform tetrahedral \tk-mesh.  Next, we offer a
refinement, test the sensitivity of the studied quantity and decide,
for each tetrahedron, whether the refinement should be accepted.
Accepting a refinement leads to the recursive generation of additional
\tk-points in a way that ensures the numerical stability of the
algorithm.

While the band structure part of the calculation uses augmented plane
waves, the Hubbard model is naturally formulated in terms of localized
orbitals.  The transformation between the two bases is accomplished by
the Wannier construction \cite{wannier_structure_1937,
  kunes_wien2wannier:_2010, marzari_maximally_1997}.

\begin{figure}
  \centering
  \includegraphics[width=\linewidth]{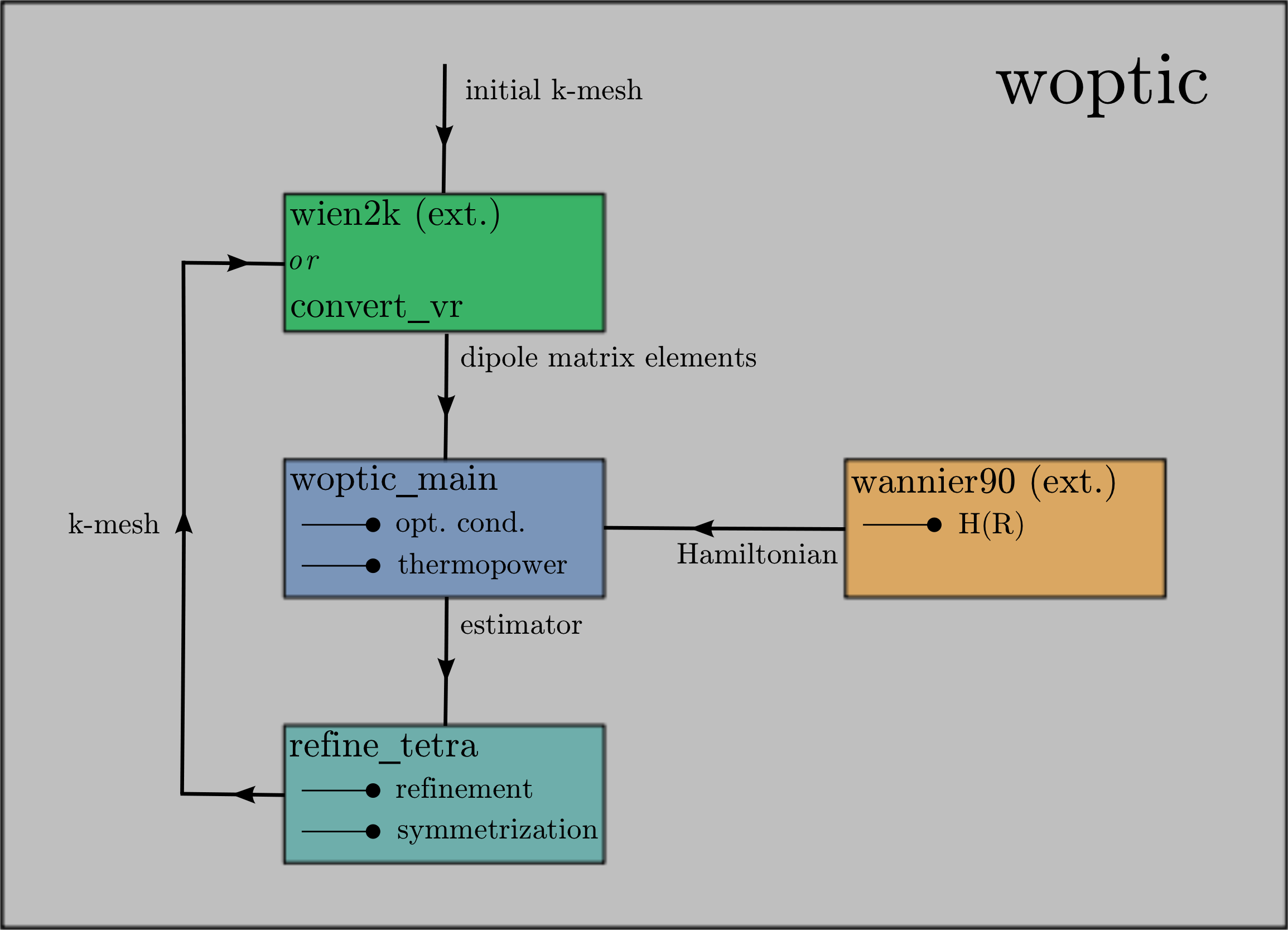}
  \caption{Schematic work flow of the \woptic algorithm with an
    adaptive tetrahedral mesh for the Brillouin zone integration.  The
    algorithm is implemented in two main programs \womain (evaluates
    the optical conductivity for a given \tk-mesh) and \reftet
    (adaptively refines the \tk-mesh) which are called by the driver
    script \woprog, together with several support programs.  \wannier
    provides the real space hopping matrix $H(\bR)$, whereas the
    dipole matrix elements are computed either by the \wien programs
    \lapwi and \optic, or by the \woptic program \convvr.  See \woptic
    user's guide for more detailed work flow diagrams.}
    \label{Fig:woptic-workflow}
\end{figure}

An overview of the work flow of the adaptive refinement program called
\woprog can be found in Fig.~\ref{Fig:woptic-workflow}.  The work flow
can be summarized as 
\begin{enumerate}
  \setcounter{enumi}{-1}
\item Choose an initial \tk-mesh and set iteration $\ell=0$~(see
  Sec.~\ref{Mesh} for the formal definition of the mesh).

\item\label{wfshort:matel} Compute the dipole matrix elements
  $v^{(\ell)}(\bk)$ of Eq.~\eqref{Eq:vk} (see Sec.~\ref{Usage} for
  details).

\item Compute the optical conductivity $\sigma^{(\ell)}$ for a given
  \tk-mesh (see Sec.~\ref{Problem} for the formula used to obtain
  $\sigma^{(\ell)}$).  Extract the information which regions of
  \tk-space have a large contribution to the integration error~(see
  Sec.~\ref{Estimation} for details on the numerical quadrature and
  how the error is estimated).

\item Stop if the change of the optical conductivity with respect to
  the previous iteration $\sigma^{(\ell-1)}$ is below a given
  tolerance,\footnote{In the current version of the code, convergence
    has to be checked manually.} %
  otherwise refine the \tk-mesh where necessary, thus obtaining a new
  \tk-mesh, and return to step \ref{wfshort:matel} with
  $\ell\gets\ell+1$ (see Sec.~\ref{Mesh} for information on the
  refinement process).
\end{enumerate}
A more detailed version of the work flow, with a description of
available modes, can be found in Sec.~\ref{Usage:details}.

This paper is structured as follows: First, in Sec.~\ref{Problem}, we
give a description of the specific numerical problem to compute the
optical conductivity.  In Sec.~\ref{Algorithm}, we specify the
tetrahedral mesh and the refinement strategy.  Furthermore, we survey
the estimation of the integration error necessary to mark tetrahedra
for refinement in Sec.~\ref{Estimation} and depict methods to increase
the numerical performance in Sec.~\ref{Performance}.  In
Sec.~\ref{Usage}, we focus on practical considerations such as the
available modes in the program and more details of the work flow and
show numerical tests.  Finally, in Sec.~\ref{Applications} we present
two applications, elementary aluminum and the vanadate \svo.

\section{Problem statement}
\label{Problem}

The Kohn-Sham Hamiltonian $H$ of \lda is diagonal in the Bloch-wave
basis.  Hence, the corresponding optical conductivity $\sigma$ can be
written in terms of the dipole matrix elements and $\delta$-functions
\cite{ambrosch-draxl_linear_2006}, setting $\hbar=1$,
\begin{multline}\label{Eq:OpticalConductivityLDA}
  \sigma_{\alpha\beta}(\Omega) =
  -\frac{e^2}{(2\pi)^2}
  \intk \sum_{c,v}
  \frac{
    \delta(\varepsilon_{c}(\bk)
    -\varepsilon_{v}(\bk)-\Omega)
  }{\Omega}
  \\
  \cdot v_{v,c}^\alpha(\bk)v_{v,c}^\beta(\bk).
\end{multline}
Here, $\Omega$ is the external frequency, $e$ is the electronic
charge, the sum is over conduction ($c$) and valence ($v$) electrons
with energy $\veps_{c/v\,\bk}$, while
\begin{equation}
  \label{Eq:vk}
  v^\alpha_{nm}(\bk) =
  -\frac\ii\me \braket{\psi_{n\bk} | \partial_\alpha | \psi_{m\bk}}
\end{equation}
are the dipole matrix elements with the Bloch states $\psi_{n\bk}$,
and $\me$ denotes the electron mass.  In general, the resulting value
for $\sigma$ depends on the number of \tk-points used in the
integration as well as the applied quadrature rule.  However, the
evaluation of the integrand in Eq.~\eqref{Eq:OpticalConductivityLDA}
is numerically cheap and one usually proceeds to uniformly refine a
given \tk-mesh until convergence.

In the following we assume that we have constructed a mapping $U(\bk)$
from $N$ Bloch states computed by \wien \cite{madsen_efficient_2001}
to $N$ maximally localized Wannier functions (\wfs) with \wannier
\cite{mostofi_wannier90:_2008}; thus we have the Hamiltonian in
Wannier space
\begin{equation}
  \label{Eq:Hk}
  H(\bk)=U^+(\bk)E(\bk)U(\bk)
  \qquad\text{with}\qquad
  E_{nm}(\bk)=\delta_{nm}\veps_{n\bk}.
\end{equation}
This means that the electronic structure is not described in terms of
bands $\veps_{n\bk}$, $1\le n\le N$, but by a Hamiltonian matrix
$H(\bk)\in\C^{N\times N}$.  This is appropriate, for example, when a
set of \wfs is required as an input for many-body calculations such as
\ldad.  Also in light of possible \ldad applications of the algorithm,
we will not consider the effects of vertex corrections.  In fact,
while they can significantly affect the \dmft results for other
response functions like the spin/charge susceptibilities
\cite{rohringer_local_2012, toschi_quantum_2012}, their contribution
to the \dmft optical conductivity is strongly suppressed (and vanishes
exactly in the single-band case) due to the \tk-structure of the
electronic current operator \cite{georges_dynamical_1996,
  tomczak_optical_2009}.  On the basis of these considerations, the
following general expression for the optical conductivity
\cite{tomczak_optical_2009, wissgott_dipole_2012} can be written
\begin{multline}\label{Eq:OpticalConductivityWoptic}
  \sigma_{\alpha\beta}(\Omega) =
  -\frac{e^2}{(2\pi)^2}
  \intkw
  \frac{f(\omega+\Omega)-f(\omega)}{\Omega}
  \cdot\\
  \Tr\left[
    v^{\W \alpha}(\bk) A(\bk,\omega+\Omega)
    v^{\W \beta} (\bk) A(\bk,\omega)
  \right].
\end{multline}
Here, $f$ is the Fermi function at a given temperature,
\begin{align}
  \label{Eq:vW}
  v^{\W \alpha}_{rs}(\bk)
  = U^+_{rn}(\bk)\, v^\alpha_{nm}(\bk)\, U_{ms}(\bk)
  = -\frac\ii\me \braket{w_{r\bk} | \partial_\alpha | w_{s\bk}}
\end{align}
the dipole matrix rotated to the basis of \tk-space \wfs
$w_{r\bk}(\br)$, and $A$ the matrix spectral function
\begin{equation}\label{Eq:GeneralizedSpectrum}
  A_{mn}(\bk,\omega)= \frac\ii{2\pi}\left[
    G_{mn}^{ret}(\bk,\omega)-G_{nm}^{ret,*}(\bk,\omega)
  \right]
\end{equation}
defined via the Green's function 
\begin{equation}\label{Eq:Greensfunction}
 G(\bk,\omega) = \big[\omega-H(\bk)-\Sigma(\bk,\omega)\big]^{-1}
\end{equation}
and the corresponding electronic self energy $\Sigma(\bk,\omega)$, all
of which are also taken to be in the Wannier basis.  Whereas in \lda,
Eq.~\eqref{Eq:OpticalConductivityLDA} scales linearly with the number
of included bands, the general version, i.e.,
Eq.~\eqref{Eq:OpticalConductivityWoptic}, scales quadratically with
the number of orbitals.  Thus, for a basis set in which $H(\bk)$ is
not diagonal, such as \wfs, it is useful to reduce the number of
\tk-space evaluations, while keeping the level of accuracy as close to
the \lda formalism as possible.

In the specific case of \ldad calculations, the $(\bk,
\omega)$-resolution required to resolve all features of the integrand
of Eq.~\eqref{Eq:OpticalConductivityWoptic} depends in particular on
the imaginary part of the self energy $\Sigma(\bk,\omega) =
\Sigma(\omega)$ contained in the \dmft Green's function
$G(\bk,\omega)$.  For small $\Im\Sigma(\omega)$, a fine $(\bk,
\omega)$-mesh is needed, since
\begin{equation}
  \label{Eq:trace}
  \Tr\left[v^\W(\bk)\, A(\bk,\omega+\Omega)\,
    v^\W(\bk)\, A(\bk,\omega)\right]
\end{equation}
becomes sharply peaked.  In many systems, the values of
$\Im\Sigma(\omega)$ vary significantly between the different orbital
manifolds, in particular when considering transitions between
localized orbitals (which have been treated, e.g., with \dmft) and
itinerant orbitals (whose description usually remains at the \lda or
at the Hartree approximation level).  Hence, a useful approach is to
adapt the \tk-mesh to the problem under consideration and use a finer
resolution only where it provides a substantial increase of accuracy.
Though the present implementation assumes local self energies
$\Sigma(\omega)$, generalization to a \tk-dependent
$\Sigma(\bk,\omega)$ is straightforward as long as
$\Sigma(\bk,\omega)$ can be obtained for any \tk-point in reciprocal
space.

\section{Algorithmic details}
\label{Algorithm}

\subsection{Tetrahedral mesh}
\label{Mesh}

\begin{figure}
  \centering
  \includegraphics[width=\linewidth]{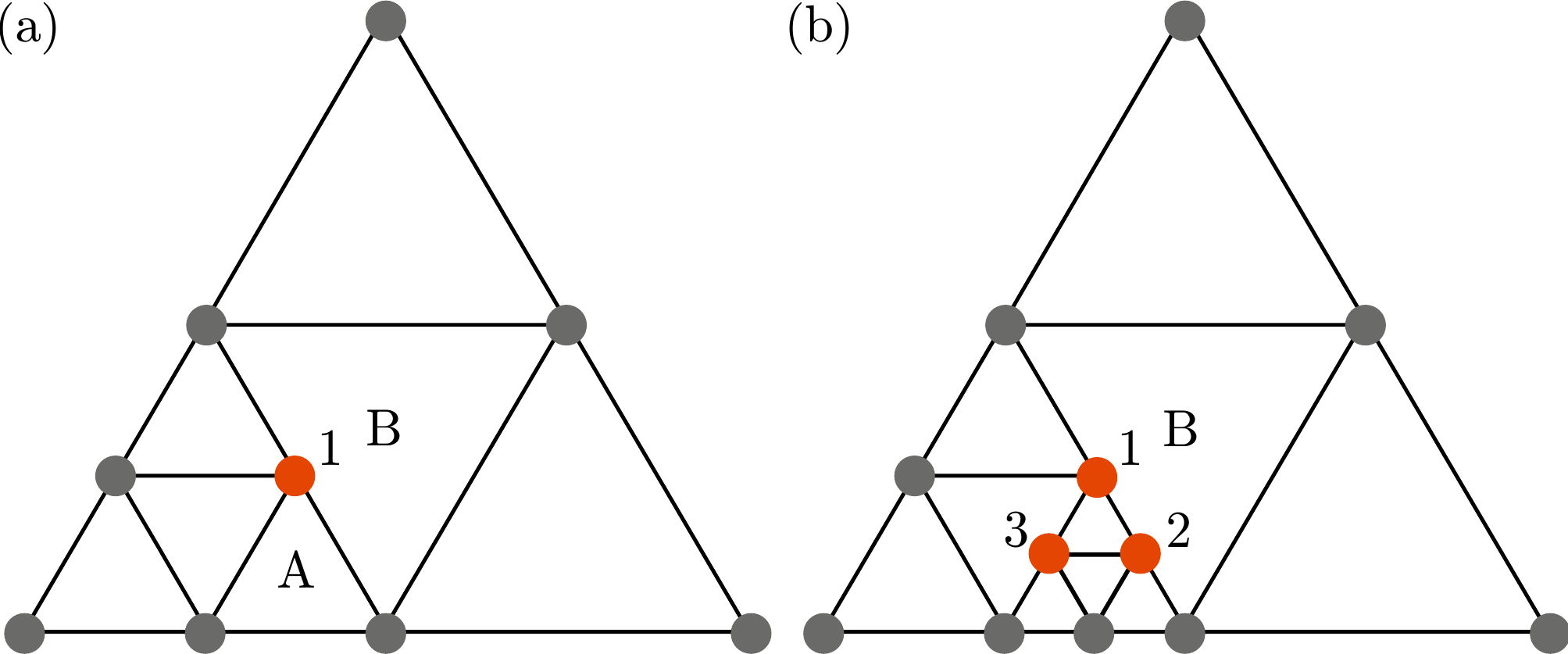}
  \caption{An example of a 2D triangulation with a hanging node (a)
    shown in red and marked $1$ (in this example we only discuss the
    nodes interior to the picture).  Upon refinement of element $A$
    shown in (a), one arrives at the triangulation (b) with two
    additional hanging nodes $2$, $3$.  This triangulation violates
    the regularity condition, since there are two hanging nodes $1$,
    $2$ on the same edge.  To make (b) regular we must also refine
    element $B$; this is the mesh closure for this triangulation.}
  \label{Fig:tetra-regularity}
\end{figure}

The non-uniform triangulation of a three-dimensional domain represents
a formidable numerical task.  The concepts surveyed in this section are
not new but combine various well-known methods, and below, we will
concentrate on the definitions necessary for the rest of this
paper.  For a more formal and complete introduction to tetrahedral
triangulation see for example Ref.~\cite{bey_tetrahedral_1995}.

We denote the set of \tk-points of a certain tetrahedral triangulation
by $\KK$ and the corresponding set of tetrahedra by $\TT$.  Then,
$N_\bk=|\KK|$ is the total number of \tk-points, and $\KK$ contains as
elements or nodes the 3D coordinates
\begin{align}\label{Eq:kmesh}
  n_m = \left[k_x\quad k_y\quad k_z\right], \qquad 1\le m\le N_\bk.
\end{align}
Furthermore, $\TT$ stores a list of vertices
\begin{align}\label{Eq:tmesh}
  T_m = \left[v_1 \quad v_2 \quad v_3 \quad v_4\right],
  \qquad 1\le m\le N_T
\end{align}
where $v_{1},\ldots,v_4\in\KK$ and $N_T=|\TT|$ is the total number of
tetrahedra (in practice, we store more information for each
tetrahedron than just the vertices, see Sec.~\ref{Estimation}).
Thus, the four vertices $v_{1},\ldots,v_4$ are nodes that define a
tetrahedron $T_m\in \TT$ and are themselves elements of $\KK$.  We
will use the term \term{vertex} only in connection to a specific
tetrahedron, while a node is a general element of the set of
\tk-points.  A special type of node is a so-called
\begin{description}
\item[hanging node:]$n_h\in\KK$ is a hanging node if it lies on an
  edge of an element $T\in\TT$ without being a vertex of $T$.
\end{description}

In the 2D visualization Fig.~\ref{Fig:tetra-regularity}(a), node 1 is
not a vertex of the central triangle B and hence a hanging node.  Of
course, a hanging node is a vertex of other triangles (here, A).

We require $\TT$ to fulfill the following two conditions:
\begin{description}
\item[regularity:] No element $T\in\TT$ has an edge with more than one
  hanging node (see Fig.~\ref{Fig:tetra-regularity}; otherwise,
  perform mesh closure, see below).
\item[shape stability:] For all tetrahedra $T\in\TT$, there is a
  predefined constant $c_S$ such that the radius of the circumscribed
  sphere $r_T$ satisfies $r_T^3/|T|\le c_S$ where $|T|$ is the volume
  of $T$.
\end{description}

A large amount of the algorithmic effort in \woptic is focused on
keeping $\TT$ shape stable and regular on refinement.  If the ratio
$r_T^3/|T|$ becomes large this indicates a highly distorted
tetrahedron (also called a degenerate element) and the numerical error
of the integration rules may become large.  A highly non-regular mesh,
on the other hand, means that nearby regions of \tk-space are resolved
very differently, see e.g.  Fig.~\ref{Fig:tetra-regularity}(b).  This
often leads to unstable convergence rates since some features of the
integrand may not be fully resolved.

\begin{figure}
  \centering
  \includegraphics[width=\linewidth]{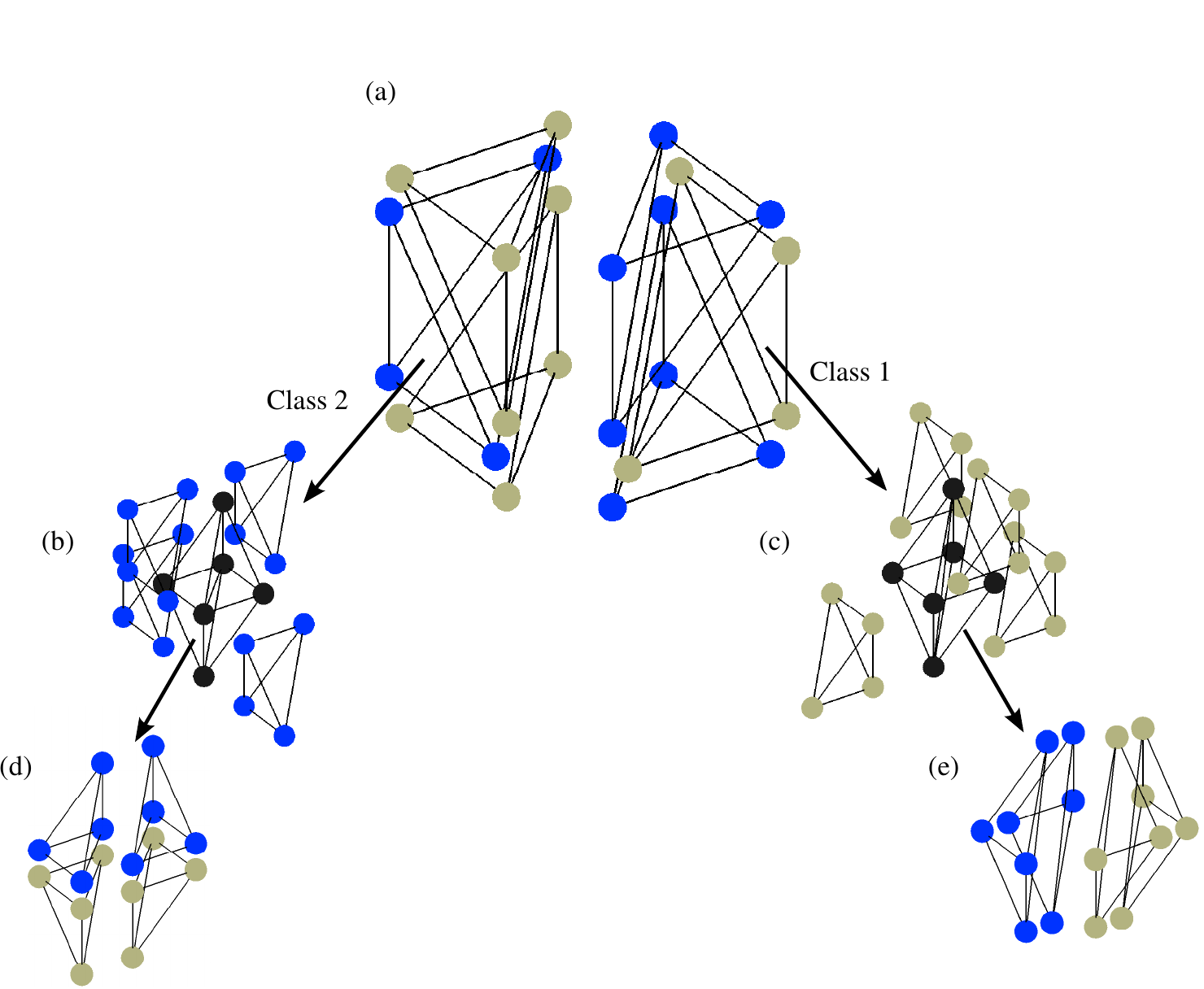}
  \caption{(a) Kuhn's refinement of a parallelepiped into eight
    tetrahedra as used by the \woptic program.  Class $1$ tetrahedra
    are marked by beige vertices while class $2$ tetrahedra are marked
    in blue.  Note that every element of class $1$ is mapped onto an
    element of class $2$ and vice versa if mirrored along the main
    diagonal plane of the cube.  The tetrahedra are first refined in
    $4$ tetrahedra similar to the original one~(having therefore the
    same class) and in a central octahedron denoted with black
    vertices, see panel (b) and (c).  Depending on the class, the
    central octahedron is further split into $4$ tetrahedra where $2$
    elements fall into the same class as the original tetrahedron and
    $2$ in the respective other class, see panel (d) and (e).}
  \label{Fig:tetra-refinement}
\end{figure}

In contrast to the 2D case of triangles, the refinement of a
tetrahedron into $8$ sub-tetrahedra of equal size is not unique and,
in general, the resulting elements cannot all be similar to the
original tetrahedron.  In the following, we will depict Ong's idea
\cite{ong_uniform_1994} for a refinement strategy where the shape of
the element is at least confined to two classes, and shape stability
is thus guaranteed.  This strategy is based on the triangulation of
the parallelepiped defined by the three reciprocal unit vectors of the
\bz into six tetrahedra of equal volume [Kuhn triangulation,
Fig.~\ref{Fig:tetra-refinement}(a)].  The resulting tetrahedra fall
into two classes, in the following denoted by class $1$ and $2$.
Specifically, it can be proven \cite{ong_uniform_1994,
  endres_octasection-based_2004} that for the refinements shown in
Fig.~\ref{Fig:tetra-refinement}, the resulting $8$ new elements of a
tetrahedron will again belong to one of the two classes.  Of the $8$
new elements, $4$ will share a vertex with the original tetrahedron
and the other $4$ will form a central octahedron.  The difference
between the two refinement methods shown in
Fig.~\ref{Fig:tetra-refinement} is how the central octahedron is split
into tetrahedra.  Since the triangulation of the central octahedron is
not unique there are other strategies to ensure shape stability,
e.g. based on the numbering of the vertices
\cite{bey_tetrahedral_1995}.  In practice, ensuring shape stability of
the mesh $\TT$ amounts to book-keeping of the classes of the
tetrahedra upon refinement.

In order to satisfy the \emph{regularity} of the mesh $\TT$ upon
refinement, we add an additional step after the standard refinement of
elements: \term{mesh closure} \cite{wissgott_space-time_2007}.
This procedure refines the elements with edges where regularity is
violated~(see Fig.~\ref{Fig:tetra-regularity}(b) for a 2D example of a
case where mesh closure is required).  This leads to neighboring
tetrahedra that will only differ by one level of refinement, since any
tetrahedron which is refined twice while all of its neighbors remain
in the initial state will automatically produce two hanging nodes on
an edge.  In this case the second refinement would trigger a
refinement of the neighboring tetrahedra as well.  Thus, when moving
through \tk-space, the refinement level changes ``smoothly'', i.e.,
regions with very fine resolution will not adjoin regions with very
coarse resolution.  These additional refinements lead to a higher
number of total tetrahedra with respect to runs without mesh closure.
Experience shows however that the price paid in performance is
acceptable, since mesh closure helps avoid unstable runs of the
algorithm.

Summarizing, we obtain the following refinement strategy, assuming we
have a regular mesh $\TT^{(\ell)}$ from the $\ell$-th iteration of
\woptic (see Fig.~\ref{Fig:woptic-workflow}) and a list of tetrahedra
marked for refinement (see next section).
\begin{enumerate}
\item Refine all marked elements of $\TT^{(\ell)}$ according to
  Fig.~\ref{Fig:tetra-refinement}(c) and (e) for class $1$ tetrahedra, and
  according to Fig.~\ref{Fig:tetra-refinement}(b) and (d) for class $2$
  tetrahedra to obtain a refined mesh $\TT^{(\ell)}_\text{ref}$.
\item Scan $\TT^{(\ell)}_\text{ref}$ for hanging nodes.
\item Mark all elements of $\TT^{(\ell)}_\text{ref}$ for refinement
  whose edges violate the regularity condition, i.e., have multiple
  hanging nodes.
\item If there are marked elements return to 1 with
  $\TT^{(\ell)}_\text{ref} \rightarrow \TT^{(\ell)}$, otherwise continue
  with $\TT^{(\ell+1)}=\TT^{(\ell)}_\text{ref}$.
\end{enumerate}
This procedure gives a tree-like nested algorithm of refinement steps,
which allows us to store the hanging nodes of
$\TT^{(\ell)}_\text{ref}$ and pass them on to the next iteration of
\woptic.  In \woptic, the refinement strategy described above is
implemented by the program \reftet (see
Fig.~\ref{Fig:woptic-workflow}).

\subsection{Integration error estimation and refinement}
\label{Estimation}

Having outlined the salient points of mesh management, let us now
discuss how elements of a mesh $\TT$ are chosen for refinement in the
first place.  Since we aim at minimizing the numerical integration
error, we have to estimate the error $\epsilon_T$ which an element
$T\in\TT$ contributes to the overall error
\begin{align}\label{Eq:OverallError}
  \epsilon_{tot} = \left\vert
    \intk g(\bk) - \sum_{T\in\TT} g_T
  \right\vert,
\end{align}
where $g(\bk)$ denotes the integrand of the optical conductivity
\begin{multline}\label{Eq:ReformulatingOC}
  \sigma_{\alpha\beta}(\Omega) =
  \intk \left(-\frac{e^2}{(2\pi)^2}\right)
  \intw \frac{f(\omega+\Omega)-f(\omega)}{\Omega}
  \cdot\\
  \Tr\left[v^\W(\bk) A(\bk,\omega+\Omega)v^\W(\bk)
    A(\bk,\omega)\right]
  \\
  \mathbin{=:} \intk g(\bk)
\end{multline}
and $g_T$ denotes an adequate tetrahedral quadrature rule~(in $g(\bk)$
we omitted the dependence on the external frequency $\Omega$ for the
moment, see below).  For the integration over the internal frequency
$\omega$, we use a straightforward summation, exploiting only the
weight factor $f(\omega+\Omega) - f(\omega)$ to limit the range of
integration.  For the \tk-integration we use two different rules:
First, a linear $4$-point rule
\begin{align}\label{Eq:4pointrule}
  g_T^\text{4p} = \frac{1}{4} \sum_{i=1}^{4} g(v_i)
\end{align}
with $v_1,\ldots,v_4$ being the vertices of $T$.  Second, rule
\eqref{Eq:4pointrule} can also be applied to the refined elements as
\begin{align}\label{Eq:32pointrule}
  g_T^\text{4pr} = \frac{1}{32} \sum_{j=1}^{8}\sum_{i=1}^{4} g(v_{ji}),
\end{align}
where $v_{ji}$ is the $i$-th vertex of the tetrahedron $T_j$
($j=1,\ldots,8$) obtained from a refinement of $T$ as introduced in
the previous section.  In the implementation, we evaluate the function
$g(\bk)$ on the $10$ points required to apply both rules
Eq.~(\ref{Eq:4pointrule}) and (\ref{Eq:32pointrule}).  The $4$
vertices of $T$ required for the rule 4p are also included in
$T_{1},\ldots,T_{8}$.  The latter additionally have $6$ midpoints on
the edges of $T$.  Thus, $T_m\in\TT$ is represented by the $10$ nodes
plus the class of the tetrahedron~(which is required for the
refinement strategy, see previous section),
\begin{equation}\label{Eq:tmesh2}
  T_m = \left[v_1\ v_2\ v_3\ v_4;\ n_{12}\ 
    n_{13}\ n_{23}\ n_{14}\ n_{23}\ n_{34};\ 
    1\text{ or }2\right],
\end{equation}
where $n_{ij}$ is the midpoint between the vertices $v_i$ and $v_j$.
Note that the nested nature of our quadrature rules allows us to
re-use values of $g(\bk)$ in following iterations of the algorithm.

To estimate the contribution which $T$ adds to the total error
$\epsilon_{tot}$, we compare the results of Eqs.~(\ref{Eq:4pointrule})
and (\ref{Eq:32pointrule}), which means that the same rule is compared
for two different levels of refinement
\cite{deuflhard_numerical_2003}.  Thus, our error estimator is
\begin{align}\label{Eq:ElementErrorEstimation}
  \epsilon_T = \left\vert\int\limits_T\! d\bk\ 
    g(\bk)-g^\text{4pr}_T\right\vert\sim
  \left\vert g_T^\text{4p}-g_T^\text{4pr}\right\vert .
\end{align}
Since the rule (\ref{Eq:32pointrule}) is obtained by applying the rule
(\ref{Eq:4pointrule}) to the sub-elements $T_{1},\ldots,T_{8}$ which
would be new elements if $T$ was refined, the error estimate
$\epsilon_T$ provides a measure of how much a refinement of $T$ would
improve the numerical integration.

The dependence of the optical conductivity
$\sigma_{\alpha\beta}(\Omega)$ on the external frequency $\Omega$ and
the directional dependence ($\alpha\beta$) have been neglected so far.
To take these dependencies into account, all error estimates of an
element are averaged,
\begin{align}\label{Eq:ElementErrorEstimation2}
  \bar{\epsilon}_T = \frac16 \frac1{N_\Omega}
  \sum_{\Omega\alpha\beta} \epsilon^{\alpha\beta}_T(\Omega),
  \qquad \alpha\beta \in \{xx,xy,xz,yy,yz,zz\}.
\end{align}
To mark certain elements for refinement, we apply a standard procedure
for adaptive mesh algorithms \cite{wissgott_space-time_2007}: an
element $T$ is marked if
\begin{align}\label{Eq:ElementErrorEstimation3}
  \bar{\epsilon}_T \ge \Theta\ \max_{T'\in\TT}\bar{\epsilon}_{T'},
\end{align}
where $\Theta\in[0,1]$ is a parameter determining the \term{harshness}
of the refinement.  A value of $\Theta=0$ means that all elements
satisfy~\eqref{Eq:ElementErrorEstimation3}, i.e. uniform refinement,
whereas large values of $\Theta$ lead to highly adaptive meshes.  In
\woptic, the error estimation is partly performed by \womain and
partly by \reftet (see Fig.~\ref{Fig:woptic-workflow}).  The former
computes the integrand, the latter calculates the error estimators
$\bar{\epsilon}_T$ and marks the elements for refinement according to
Eq.~\eqref{Eq:ElementErrorEstimation3}.

In metallic cases, the optical conductivity $\sigma(\Omega)$ for
$\Omega\rightarrow 0$ has a Drude contribution corresponding to a
Lorentzian at $\Omega=0$ which is broadened by $\Im\Sigma(0)$.  Thus,
if $\Im\Sigma(0)$ is small, the error estimator
\eqref{Eq:ElementErrorEstimation2} is often dominated by the values
around the Fermi level $\Omega=0$ and the algorithm mainly resolves
the Fermi surface.  This behavior may be adequate when one is
interested in the dc-conductivity or the thermopower, but for the
optical inter-orbital transitions at higher energies one might favor a
better description in that region.  In this case, another error
estimator instead of Eq.~\eqref{Eq:ElementErrorEstimation2} is more
appropriate:
\begin{align}\label{Eq:ElementErrorEstimation4}
  \bar{\epsilon}'_T = \frac1{6 N_\Omega}
  \sum_{\Omega\alpha\beta}\Omega\,
  \epsilon^{\alpha\beta}_T(\Omega) \qquad
  \alpha\beta \in \{xx,xy,xz,yy,yz,zz\},
\end{align}
where the additional factor $\Omega$ attributes a larger weight to the
error at higher frequencies.

\subsection{Performance and symmetry considerations}
\label{Performance}

Given that one evaluation of the function $g(\bk)$ from
Eq.~\eqref{Eq:ReformulatingOC} is numerically expensive, the number of
total evaluations should be kept as small as possible.  For this
reason, we use two techniques: ($i$) re-using the data from previous
iterations and ($ii$) taking into account the symmetries of the
crystal.  The first point is the main reason for choosing the two
nested quadrature rules Eqs.~(\ref{Eq:4pointrule})
and~(\ref{Eq:32pointrule}), since, as mentioned above, a refinement of
an element $T\in \TT$ yields at most $6$ new nodes.  Moreover,
neighboring elements with similar refinement level share nodes with
$T$.  Thus, though our integration rules are of low order, they
represent an efficient choice in terms of the number of total
evaluation points.

To understand ($ii$), i.e. how to increase the performance by
symmetry, let us define matrices $\MM \subset \R^{3\times3}$
describing the symmetry operations of the crystal in a Cartesian
coordinate system.  Furthermore, $\KK_s\subseteq\KK$ denotes the
symmetrized \tk-mesh, i.e. the reduced mesh when the symmetry
operations of $\MM$ are exploited, with the corresponding mapping
$m_s: n \mapsto n_s$ such that $n\in\KK$ and $n_s\in\KK_s$.  If one
replaces each vertex $v$ of each element of $\TT$ by its reduced
vertex $m_s(v)$, one formally obtains a new tetrahedral mesh $\TT_s$.
Note that $\TT_s$ might include elements that do not correspond to
real tetrahedra but have e.g. equal nodes when multiple vertices of an
element of $\TT$ have been mapped onto the same \tk-point in the
reduced set $\KK_s$.  After the mapping $\TT\rightarrow\TT_s$ there
are in general multiple occurrences of an element $T_s$.  For
simplicity, $\TT_s$ in the following denotes the reduced symmetrized
mesh, i.e. all elements $T_s\in\TT_s$ are only considered once and
carry a weight $w_{T_s}$, which accounts for the volume and
multiplicity of $T_s$.

The numerical quadrature to yield the optical conductivity according
to Eq.~\eqref{Eq:OverallError} is given by
\begin{align}\label{Eq:Symmetry1}
  \sigma = \sum_{T\in\TT} g_T. 
\end{align}
It is important to stress here that one cannot simply replace
$n\in\KK$ within the rule $g_T$ by $m_s(n)\in\KK_s$, since $\sigma$
and $g_T$ are tensors.  Hence, rotated quantities have to be used:
\begin{align}\label{Eq:Symmetry2}
  \sigma = \frac{1}{|\MM|}
  \sum_{W_s\in\MM}\sum_{T_s\in\TT_s}w_{T_s} W_s^+ g_{T_s} W_s
  .
\end{align}
In this approach it is sufficient to compute $g(n_s)$ for
$n_s\in\KK_s$ and this, depending on the symmetry of the problem, may
yield a considerable speed-up.

\section{Practical usage}
\label{Usage}

\begin{figure}
  \centering
  \includegraphics[width=\linewidth]{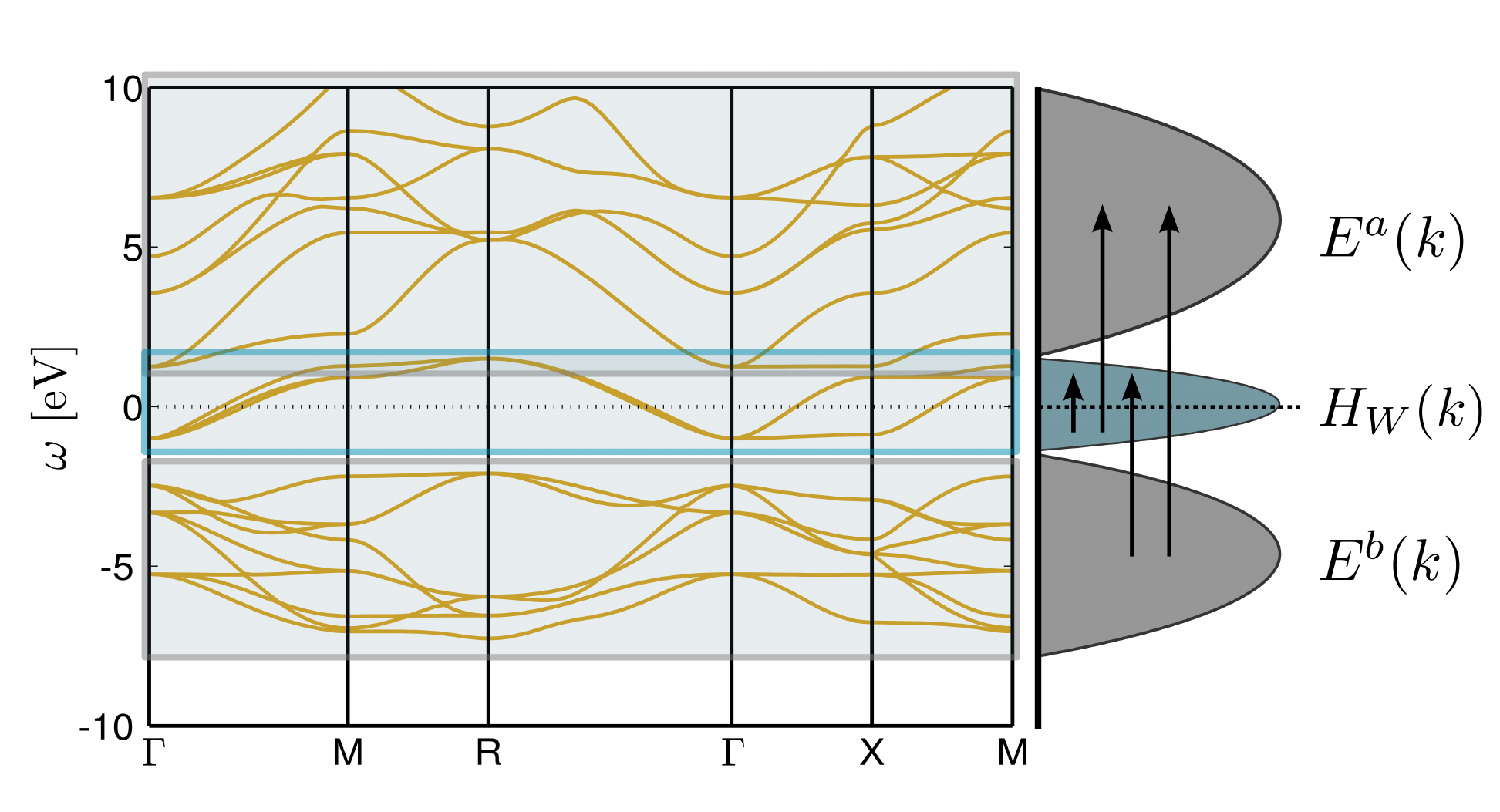}
  \caption{Possible optical transitions and corresponding Hamiltonians
    according to Eq.~\eqref{Eq:LargeHamiltonian} for various manifolds
    in a low energy model for \svo.  In the schematic visualization of
    the spectrum, the \ttg manifold where the Wannier projection is
    performed is marked $H_W(k)$ and other parts of the spectrum
    $E^{a/b}(\bk)$.}
  \label{Fig:svo-bands}
\end{figure}

After the general description of the algorithm in the previous
section, let us now turn to the connections of \woptic to the program
packages \wien and \wannier, the available modes of operation and a
more detailed work flow.  As a prerequisite to start a \woptic
calculation, two other packages are needed: \wien
\cite{madsen_efficient_2001} (including \wtow
\cite{kunes_wien2wannier:_2010}) and \wannier
\cite{mostofi_wannier90:_2008}.  This set of programs allows
constructing maximally localized \wfs from \wien (see
Refs.~\cite{kunes_wien2wannier:_2010,wissgott_transport_2012} for a
detailed description).

To employ the adaptive integration described in the previous section,
we have to be able to generate the dipole matrix $v^\W(\bk)$ and the
Hamiltonian $H(\bk)$ at arbitrary, \textit{a priori} unknown,
\tk-points.  Two approaches are implemented in \woptic: (1) matrix
elements between \wfs can be obtained by Fourier transformation from
direct space --- this is called called \Minterp mode in the program;
or (2) these matrix elements can be recomputed by \wien every
iteration --- \Moptic mode.

Two issues are central in the following:
\begin{description}
\item[\term{Gauge invariance}] of the optical conductivity
  \eqref{Eq:OpticalConductivityWoptic} and other observables, i.e. the
  independence of the random gauge of the Bloch states.  While the
  trace in Eq.~\eqref{Eq:OpticalConductivityWoptic} is gauge invariant,
  its building blocks $v^\W(\bk)$ and $A(\bk,\omega)$ are not.  It is
  therefore essential that they be expressed in the same gauge.

\item[\term{Mixed transitions,}] dipole transitions between Wannier
  states on the one hand, and Bloch states that are \emph{not}
  included in the initial Wannier projection on the other.
\end{description}

\emph{Gauge invariance} is easily ensured if approach 1 (\Minterp
mode) is available.  The Wannier construction consists in finding a
k-smooth gauge on the initial k-mesh $\KK_\W$.  This amounts to
finding a set of unitary matrices $U(\bk)$ which represent the
transformation from the initial, ``random'' gauge\footnote{More
  precisely, the gauge determined by diagonalization of $H(\bk)$ in
  the electronic structure code.} to the new gauge
\cite{marzari_maximally_1997}.  One then expresses the Kohn-Sham
Hamiltonian and the dipole matrix elements in the new gauge using
Eqs.~\eqref{Eq:Hk} and~\eqref{Eq:vW}, to repeat:
\begin{gather*}
  H(\bk) = U^+(\bk)\,E(\bk)\,U(\bk)
  \qquad \text{with } E_{nm}(\bk) = \delta_{nm}\veps_{n\bk},
  \\
  v^{\W}(\bk)=U^+(\bk)\, v(\bk)\, U(\bk).
\end{gather*}
The smoothness of the the Wannier gauge guarantees that the
corresponding basis functions (i.e. the \wfs) are exponentially
localized, and hence that the Fourier transforms of the Hamiltonian
\begin{align}
  H_{rs}(\bR) \label{Eq:HR-Fourier}
  & = \frac{1}{N_\bk}\sum_{\bk\in \KK_\W}H(\bk)\, \ee^{-\ii\bk\cdot\bR}
    = \braket{w_{r \bZ} | \hat H  | w_{s \bR}}
\intertext{ and dipole matrix elements}
  v_{rs}^\W(\bR) \label{Eq:vR-Fourier}
  & = \frac{1}{N_\bk}\sum_{\bk\in \KK_\W}v^\W(\bk)\, \ee^{-\ii\bk\cdot\bR}
    =-\frac\ii\me \braket{w_{r \bZ} | \partial_\alpha | w_{s \bR}}
\end{align}
converge rapidly with the mesh spacing of $\KK_\W$.  (Here and below,
$w_{r \bR}(\br)$ is a direct-space \wf.)

Thus we obtain direct space hopping and dipole parameters for all the
significant neighbor shells (exponentially small contributions of the
tails of the \wfs are neglected). Having the direct space
representation allows us to compute the matrix elements at an
arbitrary k-point $\ktilde$ outside $\KK_\W$ via
\begin{align}
  H(\ktilde) \label{Eq:Hk-Fourier}
  =& \sum_{\bR\in \RR_{\W}} H(\bR)\, \ee^{\ii\ktilde\cdot\bR},
  \\
  v^\W(\ktilde) \label{Eq:vk-Fourier}
  =& \sum_{\bR\in \RR_{\W}} v^\W(\bR)\, \ee^{\ii\ktilde\cdot\bR}.
\end{align}
To emphasize, the $\bR$-sums run over $\RR_\W$, the set of lattice
vectors dual to $\KK_\W$, but the localization of the \wfs allows us
to apply the reverse Fourier transform for $\ktilde \notin \KK_\W$
with excellent accuracy.  This procedure is known as \term{Wannier
  interpolation} \cite{yates_spectral_2007, wang_textitab_2006}.

\emph{Mixed transitions} arise when we want to extend the optical
conductivity, which we have written in
Eq.~(\ref{Eq:OpticalConductivityWoptic}) in terms of a trace over the
\wfs (\term{inner} or \term{Wannier window}), to include states not
covered by the Wannier projection (\term{outer} or \term{Bloch
  window}).  The motivation for the outer window is normally to
compute the optical conductivity over a larger frequency range (see
Fig.~\ref{Fig:svo-bands} for a visualization of the model and the
possible transitions).

For a \ldad calculation, the self energy $\Sigma(\omega)$, describing
correlation effects beyond \gga, has to be provided on the real axis.
In the following, this situation is referred to as \term{interacting}.
\Woptic also provides the option to set the self energy to a small
imaginary constant $\Sigma(\omega) \equiv -\ii\delta$ to mimic
broadening, e.g.  from impurity scattering.%
\footnote{This means that in the non-interacting case, the lifetime
  broadening (resulting also in a finite width of the Drude peak) is
  added ``by hand'', while in the interacting case, it arises
  naturally from \dmft.} %
We will refer to bands treated in this way as \term{non-interacting}.
By supposition, the Bloch states in the outer window are adequately
described in \lda and thus non-interacting in this sense.

When an outer window is included, we split the Hamiltonian, Wannier
transformation, and dipole matrices into Wannier, Bloch, and mixed
parts.  Likewise, the trace in the optical conductivity
\eqref{Eq:OpticalConductivityWoptic} splits into Wannier--Wannier,
Wannier--Bloch, and Bloch--Bloch terms.  Denoting the diagonal energy
matrix connected to the states above (below) the \wfs in energy by
$E^{a\, (b)}(\bk)$, we can write the ``large'' Hamiltonian, which is
block-diagonal,
\begin{align}\label{Eq:LargeHamiltonian}
  \HH(\bk)
  &= \begin{pmatrix}
    E^b(\bk) &        &           \\
             & H(\bk) &           \\
             &        & E^a(\bk)  \\
  \end{pmatrix}
  .
  \intertext{Analogously, the large Wannier transformation matrix is}
  \label{Eq:LargeUMatrix}
  \UU(\bk)
  &= \begin{pmatrix}
    \one &        &       \\
         & U(\bk) &       \\
         &        & \one  \\
  \end{pmatrix}
  .
  \intertext{It affects only the inner window and leaves the outer
    window unchanged.  Additionally, we have the large dipole matrix
    $\VV^\alpha_{mn}(\bk)=-\frac\ii\me
    \braket{\psi_{m\bk}|\partial_\alpha|\psi_{n\bk}}$ with the indices
    $n,m$ running now over all bands in the outer window instead of
    only the inner window.  Inserting $\HH(\bk)$ into
    Eqs.~\eqref{Eq:Greensfunction} and \eqref{Eq:GeneralizedSpectrum}
    yields the block-diagonal matrix spectral function $\AA(\bk,
    \omega)$ of the large system.  Together with the large dipole
    matrix in the Wannier basis}
    \VV^\W(\bk) &= \UU^+(\bk)\, \VV(\bk)\, \UU(\bk) \notag \\
    &= \begin{pmatrix}
      (v_{ij}^\alpha) & (v_{rj}^\U) & (v_{kj}^\alpha) \\[3pt]
      (v_{is}^\U)     & (v_{rs}^\W) & (v_{ks}^\U)     \\[3pt]
      (v_{il}^\alpha) & (v_{rl}^\U) & (v_{kl}^\alpha) \\[3pt]
    \end{pmatrix},
  \label{Eq:VBloch}
\end{align}
this spectral function can be used in
Eq.~\eqref{Eq:OpticalConductivityWoptic} to give a more complete
description of optical transitions in the system.

Approach 1 (\Minterp) is straightforward for the Wannier--Wannier
transitions, but not directly applicable to the mixed dipole
matrix elements
\begin{equation}
  v^\U_{ri}(\bk) = \sum_n U^+_{rn}(\bk)\,
  v^\alpha_{ni}(\bk) =
  -\frac\ii\me \braket{w_{r \bk} | \partial_\alpha | \psi_{i \bk}}
\end{equation}
because their Fourier transform $v^\U_{ri}(\bR)$
does not decay with $|\bR|$.  To salvage Wannier interpolation in the
presence of an outer window, we define the quantity
\begin{equation}
  \label{Eq:wk}
  w^{\alpha\beta}_{rs}(\bk, \omega)
  = \sum_i v^{\U \alpha}_{ri} A^{a,b}_{ii}(\bk, \omega)
  v^{\U \beta}_{is}
\end{equation}
where the index $i$ runs over all non-Wannier states (which are
non-interacting, hence their matrix spectral function $A^{a,b}$ is
diagonal).  Its Fourier transform $w^{\alpha\beta}(\bR, \omega)$
decays with $|\bR|$, albeit more slowly than $v^\W(\bR)$
\cite{assmann_spectral_2015}.

With these two interpolated quantities, and using the \wien programs
\lapwi and \optic \cite{ambrosch-draxl_linear_2006} to compute the
Bloch energies $E^{a,b}(\ktilde)$ and Bloch--Bloch dipole matrix
elements $v^\alpha_{ij}(\ktilde)$, we can evaluate the trace
\eqref{Eq:trace} at any new \tk-point.  On the other hand, the
interpolation errors from $w^{\alpha\beta}(\ktilde, \omega)$ may get
large, see next section and Ref.~\cite{assmann_spectral_2015} for
tests.  (Interpolation errors from $v^\W(\ktilde)$ are insignificant
so long as properly localized \wfs are found.)

As an alternative in the mixed case, we turn to approach 2 (\Moptic
mode): computing $\VV(\ktilde)$ and $E^{a,b}(\ktilde)$ at new
\tk-points using \lapwi and \optic.  (The Hamiltonian $H(\ktilde)$ is
still computed using Eq.~\eqref{Eq:Hk-Fourier}.)  Because the
(inner-window) spectral function $A(\ktilde, \omega)$ is computed in
the Wannier basis, but \optic yields the dipole matrix elements in the
Bloch basis, we need the Wannier transformation $U(\ktilde)$ on the
new \tk-points to mediate between them.  Since the Bloch basis is the
one in which the Hamiltonian is diagonal, we can obtain $U(\ktilde)$
by diagonalizing $H(\bk)$ (inverting Eq.~\eqref{Eq:Hk}).

The problem with approach 2 (\Moptic) lies in the arbitrariness of the
Bloch gauge.  Since the $U(\ktilde)$ obtained via Wannier
interpolation are computed by diagonalization, they are determined
only up to the phases of the respective eigenvectors.  In the
non-interacting case, where the matrix spectral functions are
diagonal, these phases evidently cancel in the trace \eqref{Eq:trace};
in fact, this reasoning can be extended to the interacting case as
long as the Wannier self energy is a scalar (diagonal in and
independent of the orbital index), e.g., because of crystal symmetry.

From a different point of view, the Bloch states $\psi_{n\bk}$,
obtained as solutions of independent eigenproblems at each \tk-point,
carry ``random'' phases.  The original $U(\bk)$ from \wannier take
these phases into account in constructing smooth functions
$w_{r\bk}(\br) = \sum_n U_{nr}(\bk)\, \psi_{n\bk}(\br)$ of $\bk$.
These phases are included both in $U(\bk)$ and $v^\alpha(\bk)$ and
hence cancel when calculating the dipole matrix elements in Wannier
space using Eq.~\eqref{Eq:VBloch}.

However, if the adaptive \tk-mesh algorithm now selects a new point
$\ktilde$, the phase of $\psi_{n\ktilde}$ is included in the
recalculated $v^\alpha(\ktilde)$ but not in $U(\ktilde)$ obtained as
explained above.  Hence, the ``random'' phase may enter into the trace
\eqref{Eq:trace} both through the Wannier--Wannier and the mixed
transitions (for the Bloch--Bloch transitions it cancels).  The
resulting \term{random-gauge problem} leads to errors in the results
whose magnitude is \textit{a priori} unknown.

So far we have assumed that the transformation between the Bloch and
Wannier states at each \tk-point is accomplished by a unitary matrix
$U(\bk)$.  This excludes the \term{disentanglement} procedure
\cite{souza_maximally_2001} implemented in \wannier, where
additionally a rectangular matrix $V(\bk)$ intervenes.  In fact,
\woptic supports disentanglement only in \Minterp mode without an
outer window (Wannier--Wannier transitions only).  It is not clear how
the method may be extended to the general disentangled case
\cite{assmann_spectral_2015}.

\subsection{Benchmarks of interpolation and random-gauge errors}
\label{optic-interp}

\begin{figure}[tb]
  \centering
  \includegraphics[width=\linewidth]{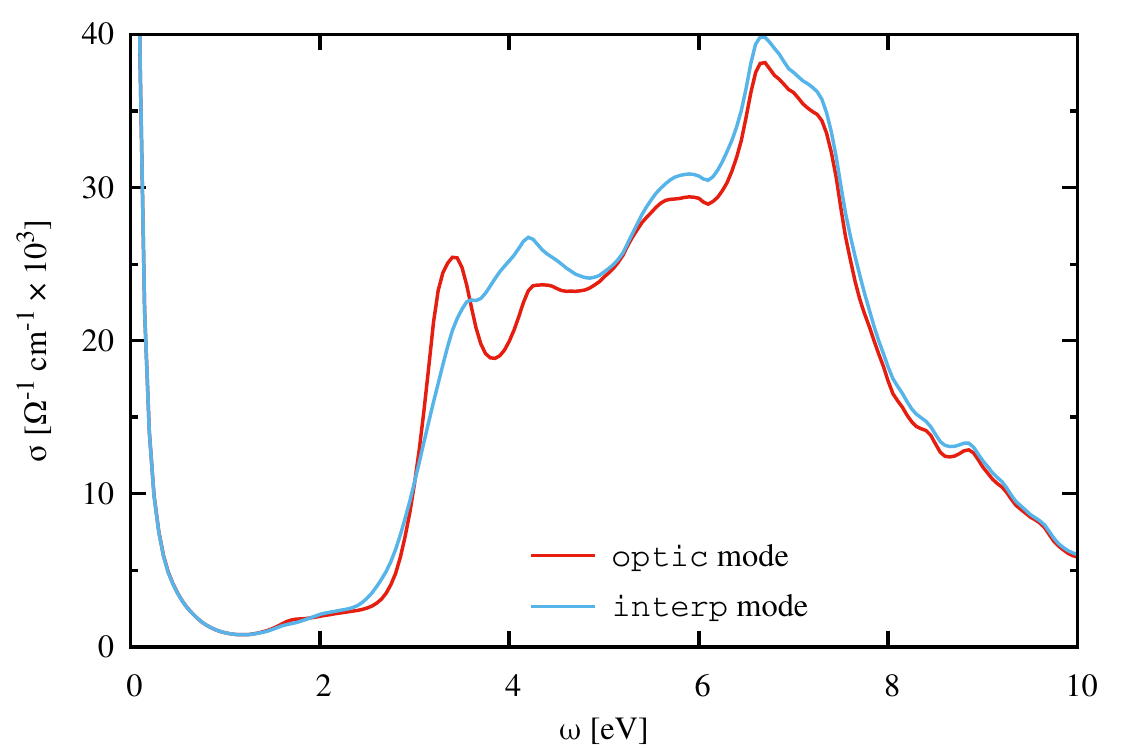}
  \caption{Optical conductivity of \svo computed in the \Moptic (red
    curve) and \Minterp (blue curve) modes.  For this model, whose
    self energy is orbital-independent due to the cubic symmetry,
    \Moptic mode can be considered exact.  Thus the difference between
    the curves reflects the interpolation errors afflicting the mixed
    (Wannier--Bloch) transitions in \Minterp mode.  Note that the
    outer window of included bands is smaller here than in
    Fig.~\ref{Fig:svo-optcond}, which is why the optical conductivity
    starts to drop off above $\omega \approx \SI{7}{\eV}$.  Here, the
    Wannier projection comprises the 3 V-\ttg bands and 14 bands are
    included in the outer window, corresponding to O-p and V-d
    states.}
  \label{Fig:optic-interp-3band}
\end{figure}

In Fig.~\ref{Fig:optic-interp-3band}, we compare the optical
conductivity of \svo from the \Moptic and \Minterp modes, including a
self energy from \dmft on the V-\ttg states (see Sec.~\ref{SrVO3} for
details on the \dmft calculation).  Since in this material the self
energy is orbital independent by symmetry, there is no random-gauge
problem, and this case can be regarded as a test for the Wannier
interpolation of $v^\W$ and $w^{\alpha\beta}$.

\begin{figure}[tb]
  \centering
  \includegraphics[width=\linewidth]{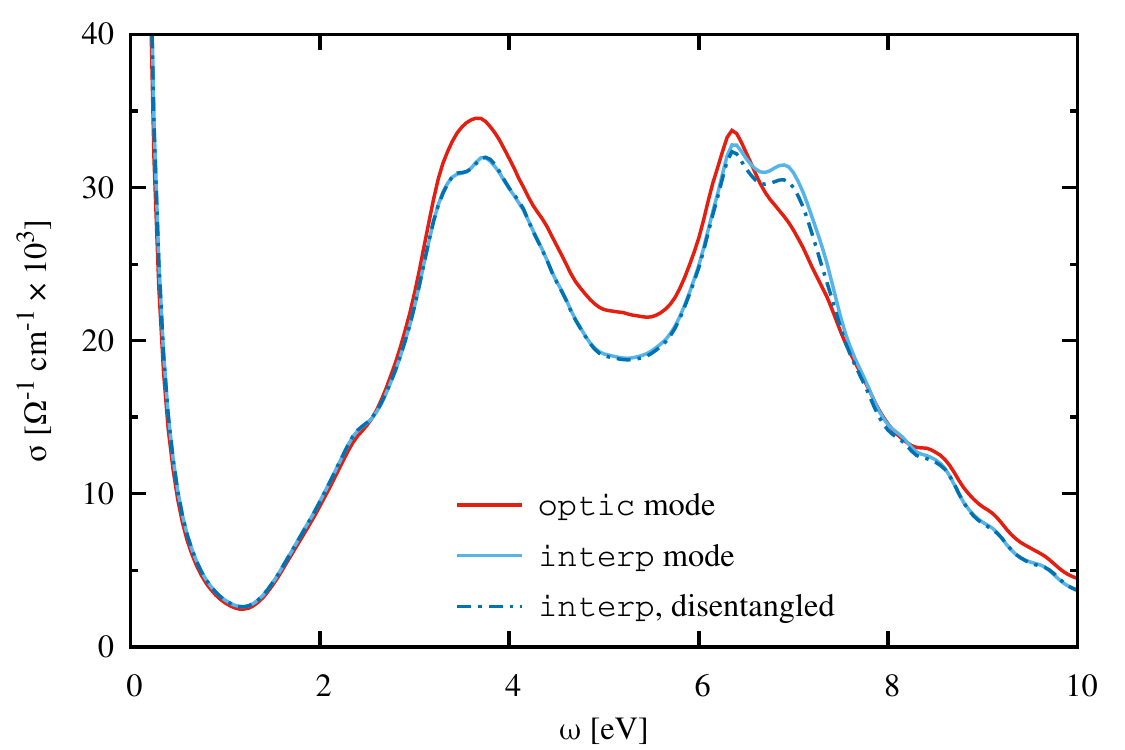}
  \caption{Optical conductivity of a model derived from \svo by
    imposing the self energy from the 3-band model on the \ttg-like
    orbitals in a \emph{14-band} Wannier projection, which includes
    the same bands as in Fig.~\ref{Fig:optic-interp-3band}: O-p,
    V-\ttg, and V-\eg.  For this model without mixed transitions,
    \Minterp mode can be considered exact, thus the difference
    reflects the random-gauge errors in \Moptic mode.  The dashed blue
    curve is from \Minterp mode using a Wannier projection with
    disentanglement of the Sr-s bands crossing the V-\eg bands at the
    $R$ point of the Brillouin zone \cite{assmann_spectral_2015}.
    Since this corresponds to a somewhat different model, complete
    agreement with the solid blue curve cannot be expected.}
  \label{Fig:optic-interp-14band}
\end{figure}

In order to quantify the random-gauge errors, we require a test case
which is complementary to the one of
Fig.~\ref{Fig:optic-interp-3band}, i.e., where interpolation is
reliable, but the gauge problem is in effect.  To this end, we
construct a Wannier projection covering exactly those bands of \svo
included in Fig.~\ref{Fig:optic-interp-3band}, and apply the V-\ttg
self-energy from Fig.~\ref{Fig:optic-interp-3band} to the \ttg-derived
orbitals of this larger projection.\footnote{This is unphysical but
  yields a convenient test case.}  Since all included transitions are
now between \wfs, we need to interpolate only $v^\W$, and can rely on
\Minterp mode as a reference; but since we apply a non-trivial self
energy only to the \ttg orbitals, $\Sigma_i(\omega)$ is now strongly
orbital-dependent and \Moptic mode suffers from the gauge problem.
The results are shown in Fig.~\ref{Fig:optic-interp-14band}.

Comparing Figs.~\ref{Fig:optic-interp-3band} and
\ref{Fig:optic-interp-14band}, we find that the errors from
$w^{\alpha\beta}$ interpolation and from the random-gauge problem are
similar in magnitude.  In both cases, the qualitative features are
preserved.  The quantitative differences must be viewed in relation to
other sources of uncertainty in these calculations.  Most importantly,
in \dmft the self energy on the real-$\omega$ axis is typically
obtained through analytic continuation from the imaginary-$\omega$
axis.  This leads to uncertainties which can easily be comparable to
the errors observed in Figs.~\ref{Fig:optic-interp-3band} and
\ref{Fig:optic-interp-14band}.  It also bears mentioning that the
orbital symmetry which protects \Moptic mode from the random-gauge
problem is broken especially sharply in the 14-band model used above
(nontrivial $\Sigma(\omega)$ on the \ttg orbitals, $\Sigma_i =
-\ii\delta$ on the others).

Fig.~\ref{Fig:optic-interp-14band} contains a third curve, which
corresponds to \Minterp mode on a model constructed with
disentanglement.  At the $R$ point of the \bz, the V-\eg bands are
entangled with Sr-s bands, as seen in the band structure of
Fig.~\ref{Fig:svo-bands}.  These unwanted bands can be removed using
disentanglement \cite{souza_maximally_2001}.  The corresponding curve
in Fig.~\ref{Fig:optic-interp-14band} is practically identical to the
one without disentanglement, except for the region around \SI{7}{\eV},
where transitions involving the entangled states are relevant.

For details on the issues discussed in the preceding paragraphs (to
wit: the random-gauge problem in \Moptic mode, interpolation of
$v^\W$ and $w^{\alpha\beta}$, and disentanglement in relation to
\woptic), we refer to Ref.~\cite{assmann_spectral_2015}.

\subsection{Program details}
\label{Usage:details}

To summarize, \woptic offers two main choices for the \term{matrix
  element mode} (corresponding to the program parameter
\code{matelmode}).  This choice determines the method to compute the
dipole matrix elements $v(\bk)$.  The modes and corresponding keywords
in the \woprog input file are:
\begin{description}
  \sloppy
\item[Wannier interpolated dipole matrix elements \code{(\Minterp)}:]
  Apply Wannier interpolation \eqref{Eq:vk-Fourier} to the dipole
  matrix elements $v^{\W \alpha}_{rs}(\bk)$ in the Wannier gauge
  \eqref{Eq:vW}, as well as to the Hamiltonian \eqref{Eq:Hk}.  In the
  presence of an outer window, the mixed transitions are calculated
  from the quantity $w_{rs}(\bk, \omega)$ \eqref{Eq:wk}, which is
  likewise interpolated \cite{assmann_spectral_2015}.

\item[\textit{Ab initio} dipole matrix elements \code{(\Moptic)}:]
  Obtain the dipole matrix elements $v^\alpha_{ij}(\ktilde)$ in the
  Bloch gauge \eqref{Eq:vk} from the \wien programs \lapwi and \optic;
  the Hamiltonian $H(\ktilde)$ from Wannier interpolation; and the
  transformation $U(\ktilde)$ to the Wannier gauge by diagonalization
  of $H(\ktilde)$.
\end{description}
\MInterp mode is reliable for the Wannier--Wannier and Bloch--Bloch
transitions, but the Wannier--Bloch terms acquire interpolation
errors.  \MOptic mode is reliable whenever the self-energy in the
Wannier gauge is diagonal and orbital independent (by symmetry, or in
the non-interacting case), but the Wannier--Wannier and Wannier--Bloch
terms acquire errors due to the random-gauge problem when it is not.
In our tests, the errors from these two issues are comparable.

We conclude this section with a summary of the detailed work flow of
\woptic.
\begin{enumerate}
  \setcounter{enumi}{-1}
\item A preliminary run of \reftet prepares the initial \tk-mesh
  $\KK^{(0)}$ and the initial set of tetrahedra $\TT^{(0)}$.  In
  \Minterp mode, obtain $v^\W(\bR)$ and $w^{\alpha\beta}(\bR, \omega)$
  (if mixed transitions are requested) from Eqs.~\eqref{Eq:vW} \&
  \eqref{Eq:wk}.  Set $\ell=0$.
\item\label{wflong:newk} Determine which \tk-points of $\KK^{(\ell)}$
  were not in $\KK^{(\ell-1)}$, i.e., find $\KK^{(\ell)} \setminus
  \KK^{(\ell-1)}$.
\item Obtain the Hamiltonian $H(\bk)$ for all \tk-points in
  $\KK^{(\ell)}\setminus\KK^{(\ell-1)}$ via Eq.~\eqref{Eq:Hk-Fourier}.
\item\label{wflong:matel} Obtain the dipole matrix $v_{mn}(\bk)$ for
  all \tk-points in $\KK^{(\ell)}\setminus\KK^{(\ell-1)}$, from the
  \wien programs \lapwi and \optic in \Moptic mode, or from
  Eq.~(\ref{Eq:vk-Fourier}) in \Minterp mode.
\item Call \womain
  \begin{enumerate}
  \item Rotate $v(\bk)$ to the Wannier basis (Eq.~\eqref{Eq:vW}).
  \item Load the self energy $\Sigma(\omega)$ and determine the
    Green's function $G(\bk,\omega)$ according to
    Eq.~\eqref{Eq:Greensfunction} for all \tk-points in
    $\KK^{(\ell)}\setminus\KK^{(\ell-1)}$.
  \item Evaluate the contributions to the optical conductivity
    $g(\bk)$ from Eq.~\eqref{Eq:ReformulatingOC} for all \tk-points in
    $\KK^{(\ell)}\setminus\KK^{(\ell-1)}$ and load the old data
    $g(\bk)$ for $\KK^{(\ell-1)}$.
  \item Perform tetrahedral integration for $\TT^{(\ell)}$ using
    Eq. \eqref{Eq:Symmetry2} and obtain the optical conductivity
    $\sigma^{(\ell)}(\omega)$
  \end{enumerate}
\item  Call \reftet
  \begin{enumerate}
  \item Determine the error estimators $\bar{\epsilon}_T$ for
    $T\in\TT^{(\ell)}$ via Eqs.~(\ref{Eq:4pointrule} \&
    \ref{Eq:32pointrule}) and (\ref{Eq:ElementErrorEstimation} \&
    \ref{Eq:ElementErrorEstimation2}), respectively.
  \item Mark the elements of $\TT^{(\ell)}$ for refinement if they
    satisfy the criterion \eqref{Eq:ElementErrorEstimation3},
    obtaining $\TT^{(\ell)}_m\subseteq\TT^{(\ell)}$.
  \item If $\TT^{(\ell)}_m$ is empty, i.e. no elements have been
    marked, set $\TT^{(\ell+1)}=\TT^{(\ell)}$ (and
    $\KK^{(\ell+1)}=\KK^{(\ell)}$) and exit from \reftet.
  \item\label{wflong:refine} Refine the marked tetrahedra
    $\TT^{(\ell)}_m$ according to their class and the rules shown in
    Fig.~\ref{Fig:tetra-refinement}, leading to the refined mesh
    $\TT^{(\ell)}_\text{ref}$ of $\TT^{(\ell)}$ and a new set of
    \tk-points $\KK^{(\ell)}_\text{ref}$.
  \item Perform mesh closure: Mark the non-refined tetrahedra of
    $\TT^{(\ell)}_\text{ref}$ for refinement if they violate the
    regularity condition, i.e. if they have more than one hanging node
    on an edge, and obtain $\widetilde{\TT}^{(\ell)}_m$.  If
    $\widetilde{\TT}^{(\ell)}_m$ is empty, the mesh is regular: set
    $\TT^{(\ell+1)}=\TT^{(\ell)}_\text{ref}$,
    $\KK^{(\ell+1)}=\KK^{(\ell)}_\text{ref}$ and exit.  Otherwise set
    $\TT^{(\ell)}_{m}=\widetilde{\TT}^{(\ell)}_m$,
    $\TT^{(\ell)}=\TT^{(\ell)}_\text{ref}$,
    $\KK^{(\ell)}=\KK^{(\ell)}_\text{ref}$ and return to step
    \ref{wflong:refine}.
  \end{enumerate}
\item If $\ell<\ell_\text{max}$ return to step \ref{wflong:newk};
  otherwise, exit.
\end{enumerate}
The key difference between the modes is in step \ref{wflong:matel},
where the matrix elements for the new \tk-points are computed.

\section{Applications}    
\label{Applications}

\subsection{Aluminum}
\label{Al}

\begin{figure*}
  \centering
  \includegraphics[width=.8\linewidth]{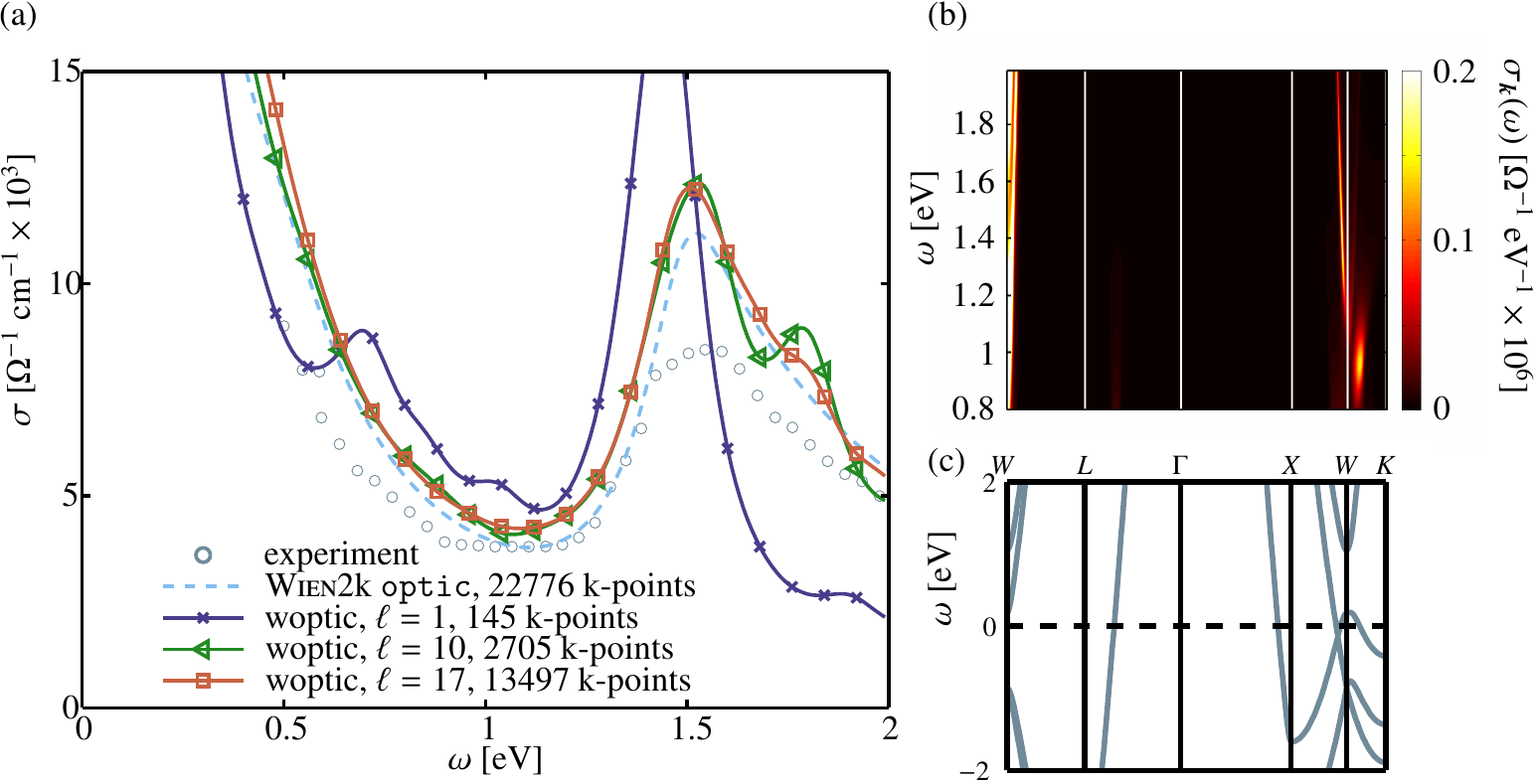}
  \caption{The non-interacting optical conductivity of \al computed by
    \woptic compared to corresponding results from the \wien \optic
    package and experimental data \cite{ehrenreich_optical_1963} in
    panel (a).  For convergence of the uniform \wien calculation a
    large number of~(symmetrized) \tk-points is required, while due to
    the adaptivity, \woptic converges for a much smaller number of
    \tk-points.  After convergence both programs yield similar
    results, in particular the experimental peak position is
    reproduced.  (An arbitrary scaling has been applied to the
    experimental results.)  The contributions to the optical
    conductivity $\sigma$ resolved in $(\bk,\omega)$-space~(b).  Only
    a small part of \tk-space around $W$ contributes to $\sigma$ and
    the contributions can be understood by identifying possible
    transitions in the band structure~(c).}
  \label{Fig:al-optcond}
\end{figure*}

\begin{figure*}
  \centering
  \includegraphics[width=.7\linewidth]{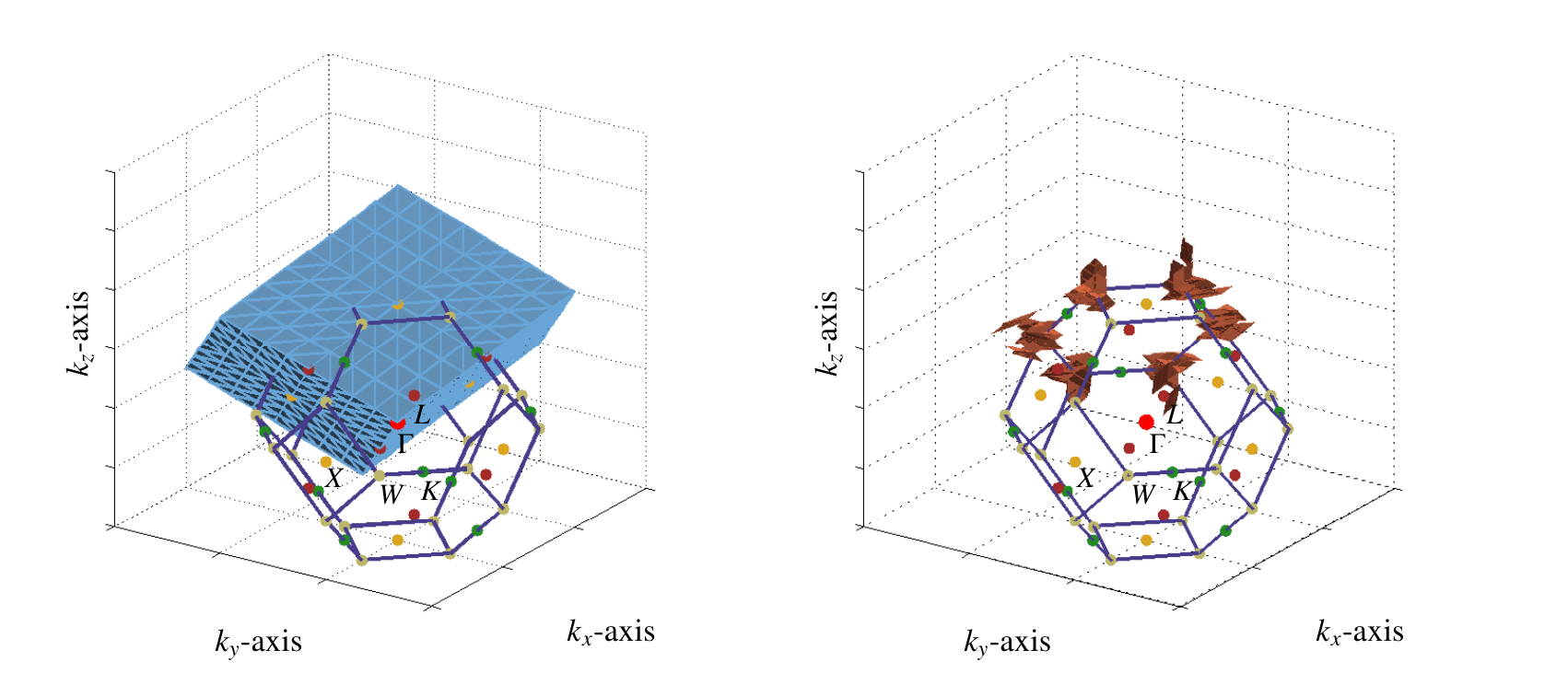}
  \caption{The unsymmetrized initial tetrahedral mesh $\TT^{(0)}$ used
    in the calculations for elementary \al with $3072$ tetrahedra and
    $4913$ \tk-points~(left).  The highly adaptive mesh after $6$
    iterations $\TT^{(6)}$ with $13152$ tetrahedra and $24667$
    \tk-points~(right).  For the sake of visualization only the
    smallest tetrahedra are shown.  The region yielding the largest
    contribution to the estimator~\eqref{Eq:ElementErrorEstimation2}
    is located close to the $W$-point around $[0.7500\quad 0.5000\quad
    0.2812]$ in terms of the primitive vectors of the reciprocal unit
    cell.}
    \label{Fig:al-kmesh}
\end{figure*}
 
As a simple example to show that our adaptive procedure can reproduce
standard \wien results at much lower computational cost we have chosen
fcc-\al.  As shown before \cite{ambrosch-draxl_linear_2006,
  lee_first-principles_1994}, there is a strong dependency of the
optical properties on the quality of the \bz integration.  Using a
regular grid and the tetrahedron method as implemented in \wien's
\optic module, up to $20\,000$ \tk-points in the irreducible wedge of
the \bz are necessary to obtain converged results for the optical
conductivity.  With our adaptive mesh the number of \tk-points can be
greatly reduced.  Note that $\sigma(\omega)$ also depends crucially on
the applied broadening scheme, and small differences may appear
between the two methods, because they are somewhat different in the
way the broadening is introduced.  The \optic module in \wien first
calculates the imaginary part of the unbroadened dielectric function,
$\epsilon_2(\omega)$, due to interband transitions and the plasma
frequency due to intraband transitions, and adds smearing later (using
Lorentzian broadening).  On the other hand, the Green's function
method uses a related broadening constant $\delta$ for the self energy
$\Sigma(\omega) \equiv -\ii\delta$ which enters into the Green's
function~\eqref{Eq:Greensfunction}.

The reason for the slow convergence of $\sigma$ with the number of
\tk-points is quite obvious when one inspects the band structure.
Consider the first interband peak at $\omega \approx \SI{1.5}{\eV}$
(Fig.~\ref{Fig:al-optcond}(a)).  In the relevant energy range, only a
narrow region of the \bz around $W$ contributes, as seen in the
\tk-resolved contributions to $\sigma(\omega)$
(Fig.~\ref{Fig:al-optcond}(b)).  Comparison with the bandstructure
(Fig.~\ref{Fig:al-optcond}(c)) shows that these contributions stem
from four bands near the Fermi level at $W$ (two particle and two hole
bands).  A regular mesh must be quite dense to properly sample these
small portions of the \bz, while our adaptive scheme is much more
efficient.  This is illustrated in Fig.~\ref{Fig:al-kmesh}, where one
can see that the regions around the $W$-point are refined most and
have the smallest tetrahedra.

\subsection{Strontium vanadate}
\label{SrVO3}

\begin{figure*}
  \centering
  \includegraphics[width=.8\linewidth]{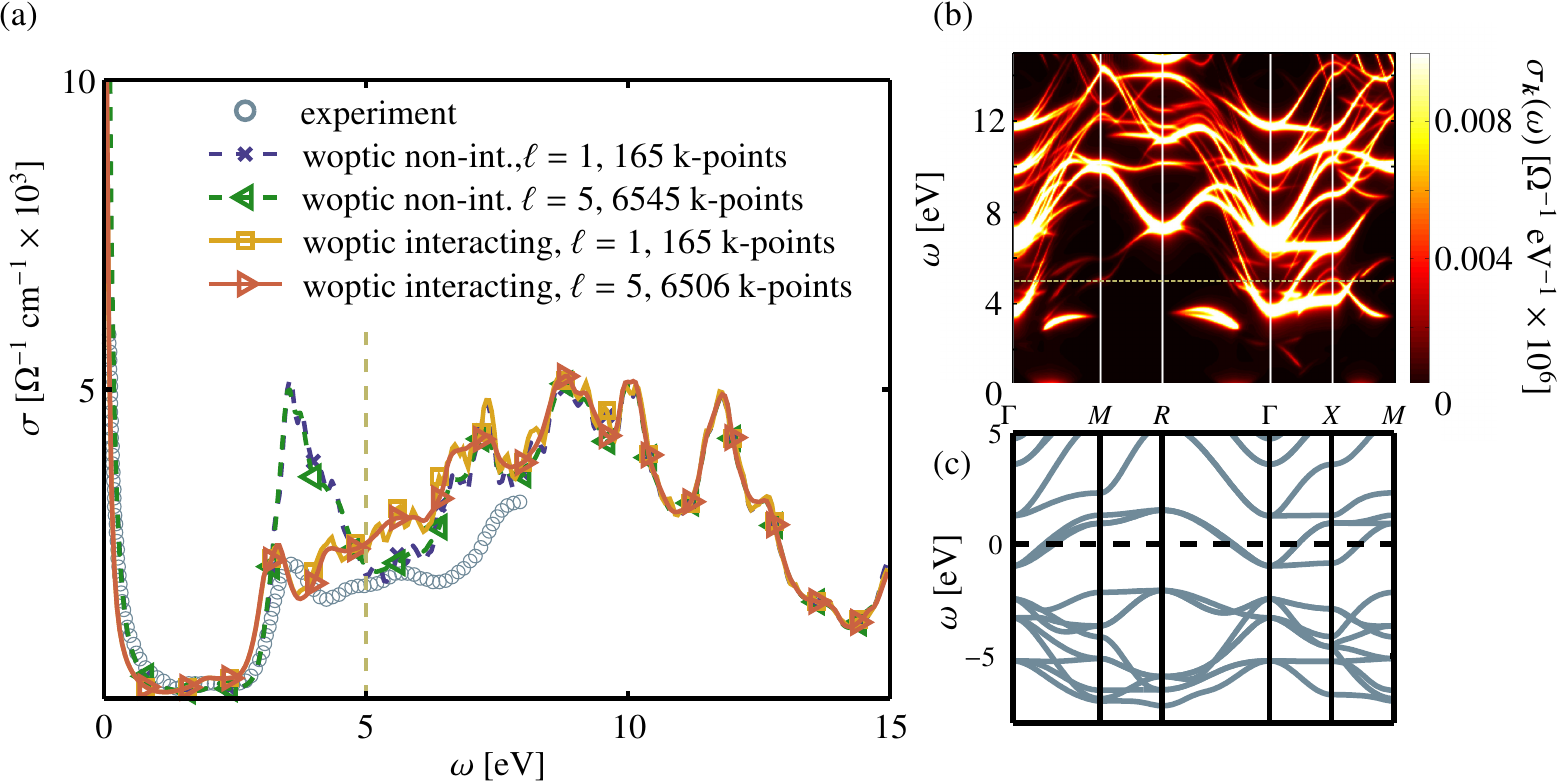}
  \caption{The interacting and non-interacting optical conductivity of
    \svo computed by the adaptive algorithm \woptic (\Moptic mode)
    compared to experiment \cite{makino_bandwidth_1998} in panel (a).
    While the interacting result is much closer to experiment, the
    convergence of the algorithm requires far fewer iterations than
    for \al both in the interacting and in the non-interacting case.
    This indicates that larger regions of \tk-space contribute to the
    optical conductivity, as also suggested by the contributions to
    $\sigma$ resolved in $(\bk,\omega)$-space (b).  (Note also the
    different scales of the contributions here and in
    Fig.~\ref{Fig:svo-kmesh}.)  A significant part of the optical
    conductivity in the energy window under investigation stems from
    O-$p\rightarrow$V-d transitions, which can be understood by
    inspection of the band structure (c).  In panels (a) and (b), the
    dashed line indicates the restricted frequency window ($0$ to
    \SI{5}{\eV}), which has been considered for the illustration of
    the evolution of the tetrahedral mesh in Fig.~\ref{Fig:svo-kmesh}.
    In the non-interacting case, the broadening, in particular the
    finite width of the Drude peak, arises solely from the small
    imaginary self energy $\Sigma = -\ii\delta$ which is added for
    this purpose.  By contrast, in the interacting case, it is a
    direct result of \dmft.}
  \label{Fig:svo-optcond}
\end{figure*}

\begin{figure*}
  \centering
  \includegraphics[width=.7\linewidth]{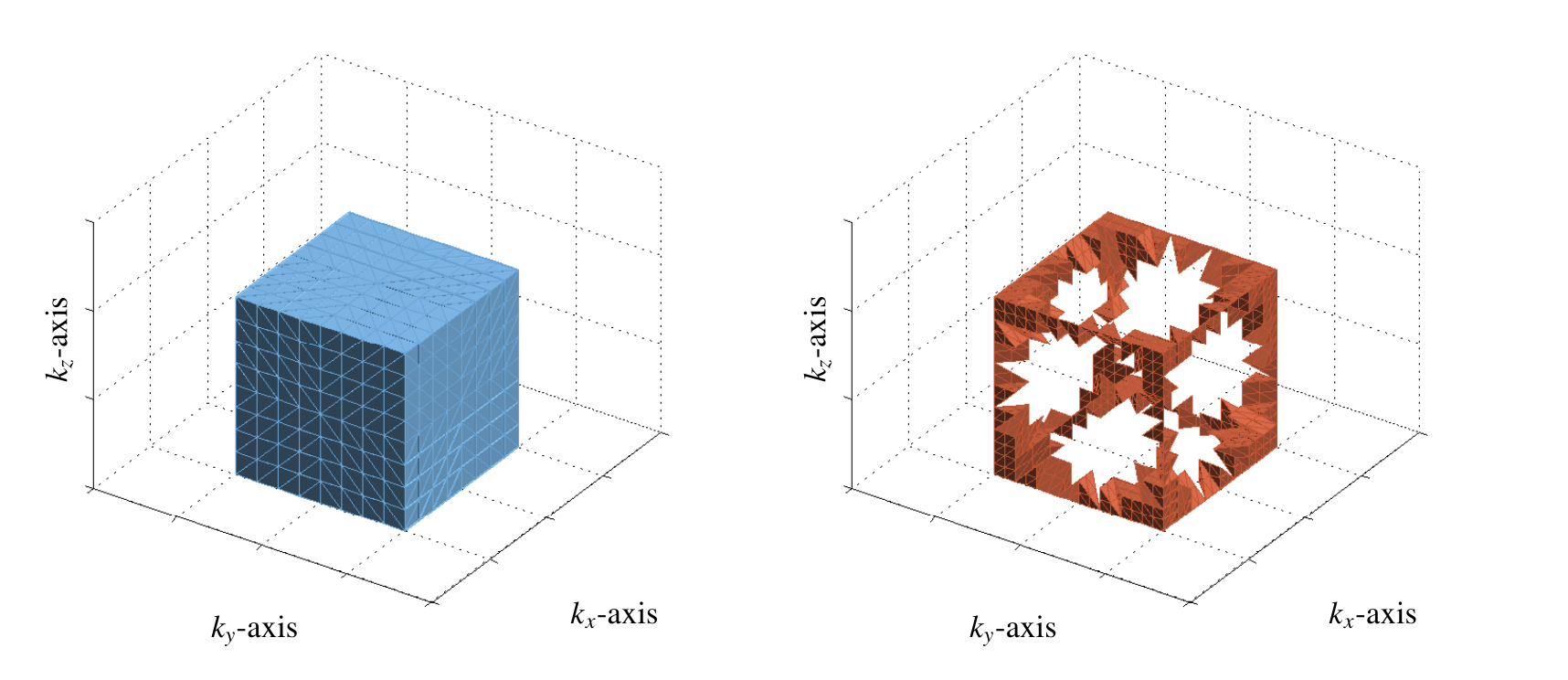}
  \caption{The unsymmetrized initial tetrahedral mesh $\TT^{(0)}$ used
    in the non-interacting calculations for \svo with $3072$
    tetrahedra and $4913$ \tk-points~(left).  After $5$ iterations, in
    the highly adaptive mesh regime ($\theta=0.9$), the mesh
    $\TT^{(5)}$ has $4920$ elements and $9533$ \tk-points~(right); for
    the sake of visualization only the smallest tetrahedra are shown)}
  \label{Fig:svo-kmesh}
\end{figure*}

As a second application of the method, we present in this section
calculations for \svo.  Its low-energy electronic structure is
dominated by the degenerate $3$d-\ttg orbitals of vanadium and it
constitutes a textbook example of a strongly correlated metal,
perfectly suited for illustrating the intermediate steps and the final
results of the \woptic package.  Specific details about the crystal
and electronic structures of \svo can be found, e.g., in
Refs. \cite{wissgott_dipole_2012, wissgott_transport_2012}.  In fact,
\svo is a good example of a situation where \lda cannot accurately
describe the low-energy optical response of the
system,\footnote{Another prototypical situation is the analysis of the
  optical spectroscopy experiments in \ce{V2O3}
  \cite{lupi_microscopic_2010, baldassarre_quasiparticle_2008,
    wissgott_dipole_2012, wissgott_transport_2012}, where the
  corrections generated by the inclusion of electronic correlation in
  \ldad are even larger than for the \svo case.} %
necessitating the inclusion of local correlation effects, e.g. by
means of \dmft.  In this work, we focus on the workings of the
adaptive algorithm rather than physical implications of our results;
see Ref. \cite{wissgott_dipole_2012} for a discussion about the
latter.

In Fig.~\ref{Fig:svo-optcond}, the intermediate ($\ell=1$) and final
($\ell=5$) results for the optical conductivity of \svo are reported,
in each case for an interacting and a non-interacting calculation.
Experimental results from Ref.~\cite{makino_bandwidth_1998} are
reproduced for comparison.  Here, the \dmft calculations have been
performed using a Kanamori interaction \cite{kanamori_electron_1963}
with parameters $U = \SI{5.05}{\eV}$, $U' = \SI{3.55}{\eV}$ and $J =
\SI{0.75}{\eV}$, consistent with the setup used in
Ref.~\cite{wissgott_dipole_2012}.  Since the random-gauge problem is
absent in this case due to the cubic symmetry (see Sec.~\ref{Usage}),
we use \Moptic mode for the interacting optical conductivity.  One
immediately notes the role of the electronic correlations, as
evidenced by a significant shift of optical spectral weight from the
Drude peak and the frequency window between $3$ and \SI{5}{\eV} to
higher energies as compared to the non-interacting calculations.  Such
many-body effects are expected due to the correlated nature of the
$3$d-\ttg orbitals of \svo, and they significantly improve the match
between theory and experiment.

A more detailed analysis of the \woptic results in
Fig.~\ref{Fig:svo-optcond} shows that the convergence of the adaptive
algorithm is much faster than for the \al case of the previous
section.  This can be understood by plotting the \tk-resolved
contributions to $\sigma(\omega)$ (i.e., the integrand of the
\tk-summation of Eq.~\ref{Eq:OpticalConductivityWoptic}), as shown in
Fig.~\ref{Fig:svo-optcond}(b): In the wide energy range considered
(i.e., up to \SI{15}{\eV}), the contributions to the optical
conductivity are spread over all of \tk-space.  As a consequence, the
adaptivity of the \woptic algorithm becomes less important and the
final adaptive mesh (not shown) essentially coincides with one
obtained in a uniform calculation.  This would be different when
focusing on the low-energy region (e.g., up to \SI{5}{\eV} as marked
by the dashed lines in Fig.~\ref{Fig:svo-optcond}) -- as is usual when
comparing with optical spectroscopic experiments.  In that case, the
predominant contribution to $\sigma(\omega)$ below \SI{5}{\eV} (apart
from the Drude peak) is from around the $\Gamma$ point, and between
$\Gamma$ and $X$.  This corresponds to the peak in $\sigma(\omega)$
located at about \SI{4}{\eV} in the non-interacting spectrum, which
originates not from optical transitions within the V-\ttg orbitals,
but rather from transitions between the O-$2p$ and the V-\eg bands.

This situation is well reflected in the evolution of the tetrahedral
mesh, reported in Fig.~\ref{Fig:svo-kmesh} for a calculation in the
window up to \SI{5}{\eV}, and performed with a highly adaptive mesh
($\theta =0.9$) for the sake of illustration.  In the right panel of
Fig.~\ref{Fig:svo-kmesh} the resulting tetrahedral mesh after $5$
iterations is visualized, showing refinements essentially from
$\Gamma\rightarrow X$, which is exactly the region we identified in
Fig.~\ref{Fig:svo-optcond}(b) and (c).  Moreover, especially along
this \tk-path, the p and the \eg orbitals mainly responsible for the
optical transitions are relatively flat and hence difficult to resolve
in $(\bk,\omega)$-space, which explains \woptic's behavior in this
case.

\section{Summary}

We have developed a flexible and efficient adaptive \bz integration
algorithm based on a recursively generated tetrahedral \tk-mesh.  In
regions where the numerical error would otherwise be large, the
\tk-mesh becomes fine, whereas it remains coarser elsewhere.  We apply
this approach to the optical conductivity in a Wannier basis, with the
possibility to include a many-body self energy $\Sigma(\omega)$ on top
of the \lda band structure.  The peakedness of the contributions in
\tk-space is determined mainly by the band structure and by the
imaginary part of the self energy, which broadens the peaks.  Thus,
weakly interacting materials, where $\Im\Sigma$ is small, tend to have
more sharply peaked contributions.

Results for \al and \svo illustrate the algorithm and its performance.
These calculations would require much more computational effort using
uniform \tk-grids.

The \textit{\woptic} package, our implementation of the adaptive
\tk-mesh algorithm in the framework of \wien, \wannier, and \dmft, is
available at \url{http://woptic.github.io}.  In addition to the
ready-made computation of the optical conductivity, dc conductivity,
and thermopower, the \tk-mesh management code may easily be adapted to
other quantities, in particular where a conventional tetrahedron
integration is impossible or impractical.

In order to include transitions involving bands beyond the Wannier
projection, an \term{outer band window} may be defined, although this
leads to certain numerical problems in some cases (see
Sec.~\ref{Usage}).  The outer bands will be described at the \lda
level, i.e. without a self energy.  Apart from the physical,
\tk-integrated quantities, \woptic also provides tools to examine the
\tk-dependent contributions (as in Figs.~\ref{Fig:al-optcond} and
\ref{Fig:svo-optcond} (b)), which often provide valuable physical
insight.

\section*{Acknowledgments}

\noindent
P.W. thanks M. Melenk and D. Praetorius for helpful discussions.  We
acknowledge financial support from the Austrian science fund \fwf
through \sfb ViCoM F41 (P.W., P.B., K.H.) and \start project Y746
(E.A.); Vienna University of Technology through an \term{innovative
  project} grant (E.A.); research unit FOR 1346 of the German science
foundation \dfg (J.K.) and its Austrian subproject \fwf \id I597-N16
(A.T.); and the European Research Council under the European Union's
Seventh Framework Program (FP/2007-2013)/\erc through grant agreement
n.\ 306447 (E.A., K.H.).  Calculations have been done on the Vienna
Scientific Cluster~(\vsc).

\bibliographystyle{elsarticle-num-modified}
\bibliography{woptic}

\end{document}